\documentclass{emulateapj}
\slugcomment{{\sc Accepted to ApJ:} September 30, 2011}
\usepackage{graphicx}
\usepackage{natbib}
\usepackage{longtable}

 \def\ergss{erg~s$^{-1}$}

 \def\gminz{$^{0.0}(g-z)$}

\shorttitle{THE OPTX PROJECT V}
\shortauthors{Trouille et al.}

\begin{document}

\title{THE OPTX PROJECT V: IDENTIFYING DISTANT AGNS\altaffilmark{1}}

\author{L. Trouille\altaffilmark{2}, A. J. Barger\altaffilmark{3,4,5}, and C. Tremonti\altaffilmark{3}}

\altaffiltext{1}{Some of the data presented herein were obtained at the W. M. Keck Observatory, which is operated as a scientific partnership among the California Institute of Technology, the University of California, and the National Aeronautics and Space Administration. The observatory was made possible by the generous financial support of the W. M. Keck Foundation.}
\altaffiltext{2}{Center for Interdisciplinary Exploration and Research in Astrophysics (CIERA) and Department of Physics and Astronomy, 2145 Sheridan Road, Evanston, IL 60208}
\altaffiltext{3}{Department of Astronomy, University of Wisconsin-Madison, 475 N. Charter Street, Madison, WI 53706}
\altaffiltext{4}{Department of Physics and Astronomy, University of Hawaii, 2505 Correa Road, Honolulu, HI 96822}
\altaffiltext{5}{Institute for Astronomy, University of Hawaii, 2680 Woodlawn Drive, Honolulu, HI 96822}


\begin{abstract}

The Baldwin, Phillips, and Terlevich emission-line ratio diagnostic
([OIII]/H$\beta$ versus
[NII]/H$\alpha$, hereafter BPT diagram) efficiently separates galaxies
whose signal is dominated by star formation (BPT-SF) from those dominated by
AGN activity (BPT-AGN). Yet this BPT diagram is limited to $z<0.5$,
the redshift at which [NII]$\lambda6584$ leaves the optical spectral
window. Using the Sloan Digital Sky Survey
(SDSS), we construct a new diagnostic, or TBT
diagram, that is based on rest-frame $g-z$ color, 
[NeIII]$\lambda3869$, and [OII]$\lambda\lambda3726+3729$ and can be used for
galaxies out to $z<1.4$. The TBT
diagram identifies 98.7\% of the SDSS BPT-AGN as TBT-AGN and
97\% of the SDSS BPT-SF as TBT-SF. Furthermore, it identifies 97\% of
the OPTX \emph{Chandra} X-ray selected AGNs as TBT-AGN. This is in contrast
to the BPT diagram, which misidentifies 20\% of X-ray selected AGNs as
BPT-SF. We use the GOODS-N and Lockman Hole galaxy samples, with their
accompanying deep \emph{Chandra} imaging, to perform
X-ray and infrared stacking analyses to further validate our TBT-AGN
and TBT-SF selections; 
that is, we verify the dominance of AGN activity in the former and star
formation activity in the latter. Finally, we address the inclusion
of the majority of the BPT-comp (sources lying 
between the BPT-SF and BPT-AGN regimes) in our TBT-AGN regime. We find
that the stacked BPT-comp source is X-ray hard ($\langle \Gamma_{\rm
  eff}\rangle=1.0_{-0.4}^{+0.4}$) and has a high X-ray luminosity to total infrared
luminosity ratio. This suggests that, on average, the X-ray signal in BPT-comp is
dominated by obscured or low accretion rate AGN activity rather than
by star formation, supporting their inclusion in the TBT-AGN regime.

\end{abstract}



\section{INTRODUCTION}

The most commonly used optical emission-line diagnostic for separating
star-forming galaxies from type II Active Galatic Nuclei (AGNs) relies on
[OIII]$\lambda5007$/H$\beta$ versus [NII]$\lambda6584$/H$\alpha$
\citep[hereafter, BPT diagram --][]{baldwin81,veilleux87}. The basic
idea is that the emission lines in star-forming galaxies
are powered by massive stars, so there is a well-defined upper limit
on the intensities of the collisionally excited lines relative to the
recombination lines (such as H$\alpha$ or H$\beta$). In contrast, AGNs
are powered by a source of far more energetic photons, making the collisionally
excited lines more intense relative to the recombination lines. Two
demarcations are commonly used for identifying AGN-dominated galaxies
versus star formation dominated galaxies -- (1) the
\citet{kewley01} theoretical division between galaxies whose extreme
ultraviolet (EUV) ionizing radiation field is dominated by an AGN
($>50$\%) and those dominated by star formation and (2) the
\citet{kauffmann03} empirical demarcation based on the location of the Sloan
Digital Sky Survey (SDSS; \citealt{york00}) 
star-forming galaxies. Galaxies that
lie between these two curves are often referred to as composite
galaxies. Hereafter, we
refer to these categories as BPT-AGN, BPT-SF, and BPT-comp. 

The BPT diagram is limited in its use with optical spectra to galaxies
with $z<0.5$ (the redshift at which [NII] leaves the optical spectral
window). A number of groups have tried to extend
optical emission 
line diagnostics to higher redshifts by only using lines at the blue
end of the spectrum. \citet{lamareille10} replaced [NII]/H$\alpha$
with [OII]$\lambda 3726+\lambda 3729$/H$\beta$, creating the `blue
diagram', which can be used to classify galaxies out to $z<0.9$ (see
also \citealt{tresse96,
  rola97,lamareille04}). They find that this diagnostic is very successful at
identifying star-forming galaxies (with $>99$\% of the BPT-SF still
classified as SF-dominated) but that it requires complementary
diagnostics to robustly classify BPT-AGN (see \citealt{marocco11}). We also note
that [OII] and H$\beta$ are relatively distant in wavelength, requiring
more careful calibration and extinction corrections. 

Recently, \citet{yan11} and \citet{juneau11} introduced the CEx
and MEx diagnostics, respectively, in
which they replaced [NII]/H$\alpha$ with rest-frame $U-B$ color (CEx) or
with stellar mass (MEx; see also \citealt{weiner07}). Both
diagnostics recover the BPT-AGN
classification very well, with 95.7\%  and $>99$\% of the BPT-AGN still
classified as AGNs in the CEx and MEx diagnostics,
respectively. However, because these diagnostics rely on H$\beta$ at $\lambda
4861$~\AA, their use with optical spectra is limited to $z<1$. 

At higher redshifts the \citet{stasinska06} DEW diagnostic, based
on $D_n[4000]$, [NeIII]$\lambda3869$, and
[OII] (see also \citealt{rola97}), can be used with optical
spectra out to $z<1.4$, at which point the lines move into the
infrared. While these emission lines are not as strong as [OIII] and 
H$\beta$, their ratio is the only choice for pushing optical spectra to
these higher redshifts. [NeIII] emission indicates the presence
of highly ionized gas and is much stronger than [OII] in
high-excitation AGNs. However, because $D_n[4000]$ requires a
sufficiently high signal-to-noise continuum (i.e., using SDSS,  we
find that the uncertainty on 
$D_n[4000]$ ranges from 5\% at $SN/$\AA~$=5$
to 25\% at $SN/$\AA~$=1$; see also \citealt{cardiel98}) and requires
the survey to be spectrophotometrically calibrated, its usage with
distant galaxies is limited.  

In this article we examine whether rest-frame $g-z$ color, which
requires near-infrared imaging at the higher redshifts, is a
compelling replacement for $D_n[4000]$. While the ratio
of [NeIII]/[OII] alone effectively separates metal-rich
star-forming galaxies from AGNs, metal-poor star-forming galaxies have
high values of [NeIII]/[OII] (as a result of less line blanketing
enabling a harder stellar 
radiation field). Fortunately, metal-poor galaxies also tend to be
bluer (see Fig.~7 in \citealt{tremonti04}), so we can use their
color to distinguish them from AGNs (which tend to be bulge dominated
and redder, see \citealt{yan11}). We refer to this new diagnostic as
the TBT diagram. 

We first establish the reliability of the TBT diagnostic in reproducing
the BPT classifications at low redshifts using SDSS, the largest
spectroscopic sample to date of emission-line galaxies. We then
test the TBT diagnostic at higher redshifts using the highly
spectroscopically complete OPTX X-ray selected
sample of AGNs \citep{trouille08,trouille09,trouille10}. The
misidentification of X-ray selected AGNs as SF-dominated galaxies is a 
potential issue for all optical emission-line diagnostic diagrams. In
\citet{trouille10} we found that $\sim20$\% of 
the $L_X>10^{42}$~erg~s$^{-1}$~\emph{Chandra} X-ray selected AGNs in
our OPTX sample that have [OIII], H$\beta$, [NII], and H$\alpha$
fluxes with signal-to-noise greater than five are
misidentified by the BPT diagram as BPT-SF (see also
\citealt{winter10} for 
evidence of this in the \emph{Swift} BAT sample and
\citealt{bongiorno10} for evidence of this in the XMM-COSMOS
sample). Existing high-redshift 
optical emission-line diagnostics also misidentify a significant
fraction of X-ray selected AGNs, e.g., $\sim8$\% and $\sim22$\% of X-ray
selected AGNs in the MEx and CEx diagrams lie in the MEx-SF and CEx-SF
regimes, respectively.  

Stacking techniques have been widely used in X-ray astronomy to study
the average properties of source populations selected to have certain
well-defined properties and which are too X-ray faint to be detected
individually
\citep[e.g.,][]{brandt01,alexander01,hornschemeier02,nandra02,georgakakis03,laird05,lehmer05,lehmer08,treister09}. As a proof-of-concept of the TBT
diagnostic, we perform an X-ray stacking analysis of the TBT-SF and
TBT-AGN in the highly
spectroscopically complete Great 
Observatories Origins Deep Survey North (GOODS-N;
\citealt{giavalisco04}) with accompanying \emph{Chandra}
Deep Field North (CDFN; \citealt{alexander03}) imaging to determine
whether they are, on average, X-ray soft or X-ray hard. 

An X-ray hard source is indicative of obscured AGN activity
or the presence of 
high-mass X-ray binaries (HMXBs) associated with ongoing star formation. An
X-ray soft source is indicative of unobscured AGN activity
or the presence of low-mass X-ray binaries (LMXBs) associated with old stellar
populations. In order to distinguish between these scenarios, we
perform an infrared (IR) stacking analysis using 
 the \emph{Spitzer Space Telescope} 24$\mu$m data. Numerous studies
 have found a relation betwen the X-ray and 
 IR luminosities for star-forming galaxies and for AGNs 
 \citep{ptak03,persic04,alexander05,teng05,georgakakis07}. We compare
 the X-ray and IR luminosities for our stacked TBT-SF and stacked TBT-AGN
 with the results from these studies to verify the
 dominance of star formation activity in the former and AGN activity
 in the latter.  

We then perform X-ray and IR stacking analyses of the BPT categories to
confirm the presence of AGN activity in BPT-comp, as suggested by the
TBT diagnostic.  In most optical
emission-line diagnostics, the BPT-comp lie in a transition region or
within the SF-dominated regime. For example, \citet{kewley06} find
that BPT-comp are 
indistinguishable from HII regions and are significantly distinct from Seyferts in an
[OIII]/[OII] versus [OI]/H$\alpha$ plot. In both the `blue diagram' and the CEx
diagnostic, the majority of BPT-comp lie within their SF-dominated
regimes ($\sim83$\% and $\sim75$\%, respectively). In contrast, the
bulk of BPT-comp lie within our TBT-AGN regime (see also 
the MEx diagnostic). Because BPT-comp are a
significant percentage of the overall low-redshift
emission-line galaxy population  (e.g., in SDSS\footnote{Here
  we have restricted the SDSS DR8 sample to emission-line 
  galaxies whose [NII], H$\alpha$, [OIII], and H$\beta$ fluxes have a
  signal-to-noise ratio greater than five.}, BPT-SF,
BPT-comp, and BPT-AGN comprise 69\%, 20\%, and 11\%, respectively, of
the overall population), inclusion or exclusion of BPT-comp in AGN
samples can have an important impact on results. In order to have the
necessary statistics required to do a robust stacking analysis, we
combine our GOODS-N galaxy sample with galaxy samples from two Lockman
Hole (LH) fields, all of which have deep \emph{Chandra} imaging. 

The structure of the paper is as follows. In Section
\ref{sample} we briefly describe the SDSS sample, our OPTX X-ray
selected sample of AGNs, and our optical spectroscopic samples of the
GOODS-N/LH fields. In Section \ref{TBT} we use the SDSS sample to
calibrate our TBT diagnostic to match the BPT diagram. In Section
\ref{xray} we determine how well our TBT-AGN selection matches with an
X-ray selection of AGNs. In Section \ref{stack} we do X-ray and
IR stacking analyses to verify the reliability of our TBT-AGN and
TBT-SF classes. In Section \ref{stackBPT} we do X-ray and IR stacking
analyses on our BPT-comp to determine whether the implications of the
TBT diagnostic with respect to the dominance of AGN activity in BPT-comp are
confirmed. In Section \ref{disc} we compare our TBT diagnostic to
other diagnostics for separating star-forming galaxies from AGNs.
In Section \ref{summary} we summarize our results. 

All magnitudes are in the AB magnitude system. We assume
$\Omega_M=0.3, \Omega_{\Lambda}=0.7$, and H$_0=70$ km s$^{-1}$
Mpc$^{-1}$. 


\section{Sample}
\label{sample}

\subsection{SDSS: Low-Redshift Galaxy Sample}

SDSS has obtained deep, multi-color images covering more than a
quarter of the sky with follow-up spectroscopy of over a million
objects. Here we use the SDSS spectroscopic data from Data Release
8 (DR8; \citealt{aihara11}). We use the
emission-line fluxes measured by the MPA-JHU group as described in
Section 4.3 of the data release paper. These fluxes are estimated from
simultaneous Gaussian fits to the continuum subtracted spectra to
account for stellar absorption and line blending. DR8 provides spectra
for 868,492 different galaxies. The SDSS spectral range is 
$3800-9200$~\AA.  Since the [OII]
line lies at $3726$~\AA~and the [NII] line lies at 6584~\AA, we limit our SDSS
samples to galaxies with $0.02<z<0.35$. Our SDSS BPT sample consists
of the 243,865 SDSS galaxies that have H$\alpha$, [NII], H$\beta$, and [OIII]
fluxes with signal-to-noise ratio (SNR) greater than
five (out of the 818,333 spectra for different galaxies in the DR8
sample). Of these SDSS BPT galaxies, 23,048 also have
both [NeIII] and [OII] fluxes with SNR~$>5$. We refer to these as our
SDSS TBT sample. 

\subsection{OPTX: X-ray Selected Sample of AGNs}
\label{optx}

The OPTX sample consists of 1789 X-ray sources in two
intermediate depth wide-field surveys in the Lockman Hole region of
low galactic column density and one
deep pencil-beam survey 
(CDFN; \citealt{brandt01,alexander03}). The Lockman Hole fields are
the \emph{Chandra} 
Large Area Synoptic X-ray Survey (CLASXS; \citealt{yang04}) and the 
\emph{Chandra} Lockman 
Area North Survey (CLANS; \citealt{trouille08,trouille09,wilkes09}). We have
spectroscopically observed 84\% of the OPTX sources using the DEep
Imaging Multi-Object Spectrograph 
\citep[DEIMOS;][]{faber03} on the Keck II 10~m telescope and the HYDRA
multi-object spectrograph on the WIYN 3.5~m telescope (for details of the
observations and reduction process, see \citealt{trouille08}). 

In \citet{trouille08,trouille09} we used the X-ray fluxes and
spectroscopic redshifts to calculate rest-frame
$2-8~\rm keV$ luminosities, $L_X$. At $z<3$ (which is all we consider
in this paper), we calculated the luminosities from the
observed-frame $2-8~\rm keV$ fluxes, assuming an intrinsic X-ray
spectral index of
$\Gamma=1.8$.  That is, $L_X=f \times 4 \pi d_L^2
\times k-$correction, where for $z<3$, $k-$correction  $=(1+z)^{\Gamma
  -2}$ and $f=f_{2-8~\rm keV}$. Using the individual X-ray spectral
indices (e.g., $\langle 
\Gamma \rangle \sim 1.75$ with a dispersion of $\approx 0.33$ derived
by \citealt{tozzi06} for the X-ray bright CDFS sources), rather than
the universal X-ray spectral index of 
$\Gamma=1.8$ adopted here, to calculate the $k-$corrections would
result in only a small difference (an average factor of 0.9) in the
rest-frame luminosities.  We have not corrected the X-ray luminosities
for absorption since these corrections are small in the $2-8~\rm keV$
band (e.g., \citealt{barger02}), and we are only using
the X-ray luminosities to identify sources as X-ray AGNs. 

In the following, we limit our study to the 561 OPTX X-ray sources with
spectroscopic redshifts, whose $2-8~\rm
keV$ flux has a significance greater than $3~\sigma$, and whose 
$L_X>10^{42}$~\ergss~(247, 163, and 151 sources from the CLANS, CLASXS, and
CDFN fields, respectively). $L_X>10^{42}$~\ergss~is a commonly used
conservative threshold for AGN activity \citep{hornschemeier01,barger02,szokoly04,silverman05,coil09} that is based on
energetic grounds \citep{zezas98,moran99a}. Using
the calibration by \citet{ranalli03}, one
would need a star formation rate (SFR) of $200~M_{\sun}$~yr$^{-1}$ to
produce enough X-ray luminosity from non-AGNs to cross this
threshold. In \citet{trouille10} we found a $<5$\%
contamination rate of our OPTX sample by sources with SFR
$>200~M_{\sun}$~yr$^{-1}$. We determined this using the 
\citet{magnelli09} space densities at a range of redshifts for ultraluminous 
infrared galaxies (ULIRGs) with estimated SFR $>172~M_{\sun}$. 

Our study focuses on optical emission-line ratio
diagnostics based on flux ratios. We compute the relative line fluxes from the
spectra using the \citet{tremonti04} software. In brief, we
subtract the stellar continuum and absorption lines by fitting a
linear combination of single stellar population models of different
ages \citep{bruzual03}. We remove any remaining residuals from the
continuum using a sliding 250~\AA\ median. The relative line
fluxes and errors are estimated from simultaneous Gaussian fits to the continuum
subtracted spectra.  

Our DEIMOS spectra are of high quality from $\sim 4800$~\AA~to $\sim
9300$~\AA, such that the [OII] and [NeIII] lines lie within 
our spectral window from $z\sim 0.3-1.4$. Of our 561 OPTX X-ray
sources, 197 are non-broad-line AGNs
(FWHM~$<2000$~km~s$^{-1}$; hereafter non-BLAGNs) and lie within this
redshift range. We only include non-BLAGNs in this analysis, since
the narrow lines in BLAGNs are overwhelmed by the emission from the
broad-line region.  We identify 103 OPTX X-ray
selected AGNs that have [NeIII] and [OII] fluxes with SNR~$>5$. 

\subsection{GOODS-N/LH: Higher-Redshift Galaxy Sample}

Each of the three OPTX fields is
the focus of a comprehensive spectroscopic follow-up of all galaxies
in the field. The CDFN
encompasses the intensively studied GOODS-N field, which we describe
in more detail below. The CLANS and CLASXS fields reside in the
Lockman Hole (LH) region of low Galactic column density. 

A random spectroscopic sample of 3082 $K_{s,\rm AB}<21.5$ galaxies in
these two LH fields was obtained using DEIMOS on Keck~II (L. Cowie,
priv.~comm.). We hereafter refer to this combined sample as the LH
galaxy sample. 

The GOODS-N field has among the deepest
images ever obtained in a number of bandpasses, including the
\emph{Chandra} 2 Ms CDFN image. It also has been the
target of extensive spectroscopic observations over the years
\citep[e.g.,][]{cohen00,wirth04,cowie04,barger08,cooper11}. With its
high optical spectroscopic completeness and deep X-ray coverage, the
GOODS-N catalog 
provides an ideal sample for studying the average X-ray properties of
optically selected samples of AGNs. 

\citet{barger08} presented a highly complete spectroscopic survey of
the GOODS-N field. Over the
years a number of groups have made observations of 
this region, first primarily using the Low-Resolution Imaging
Spectrograph (LRIS; \citealt{oke95}) on the Keck I 10 m telescope and
later using DEIMOS. Barger et al.~added to these samples by
observing all missing or unidentified galaxies to date with DEIMOS. In order to
provide a uniform spectral database, they also re-observed sources
where the original spectra were of poor quality or where previous
redshifts were obtained with instruments other than DEIMOS. The redshift
identifications are now greater than 90\% complete to magnitudes of
F$435W_{\rm AB}=24.5$ and $K_{s,\rm AB}=21.5$ and to $24~\mu$m fluxes of 250
$\mu$Jy. The final Barger et al.~catalog provides spectroscopic
redshifts for 2710 galaxies in this field.

We compute the relative line fluxes for the relevant emission
lines in the combined GOODS-N/LH galaxy sample using the same method
and software as 
described for the OPTX X-ray sample above. To create our GOODS-N/LH
BPT diagram, we use the 727 sources that have [OIII], H$\beta$, [NII], and
H$\alpha$ fluxes with SNR~$>5$. For our GOODS-N TBT diagram, we use the 670
sources that have [NeIII] and [OII] fluxes with SNR~$>5$. As discussed
in the previous subsection, because of the DEIMOS
spectral window, the GOODS-N sources plotted in our TBT
diagnostic have a redshift range from $z=0.3-1.4$. 

\begin{figure}[!htp]
\epsscale{1.2}
\plotone{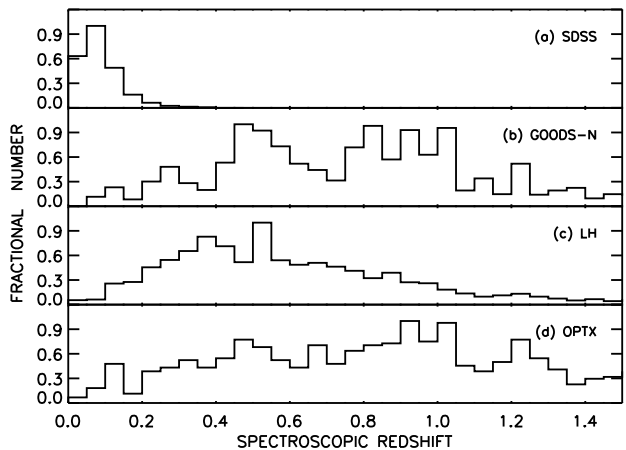}  
\caption{Spectroscopic redshift distribution out to $z=1.5$ for (a) the SDSS BPT
  sample, (b) the GOODS-N galaxy survey, (c) the 
  LH galaxy survey, and (d) the OPTX survey. The different redshift
  ranges seen in (a)-(c) reflect the magnitude limits of the different surveys.}
\label{zhist}
\end{figure}

In Figure \ref{zhist} we show the redshift distributions (out to
$z=1.5$) for the SDSS, GOODS-N, LH, and OPTX
samples. The large SDSS sample has a median redshift of $\langle z
\rangle \sim 0.1$, whereas the GOODS-N, LH, and OPTX surveys
allow us to test our TBT diagnostic out to $z=1.4$. 


\subsection{Rest-frame $g,z$ Magnitudes}

We have $u,g,r,i,z$ magnitudes for the SDSS galaxies \citep{abazajian09},
$g,r,i,z,J,$\\$H,K_s$ magnitudes for the OPTX AGNs and LH galaxies
\citep{trouille08,keenan10}, and
$B,V,R,I,Z,J,H,K_s$ magnitudes for the GOODS-N galaxies
\citep{giavalisco04,barger08,keenan10}. We transform the observed
photometry into $g$ and $z$ magnitudes at $z=0$ using kcorrect v4\_1\_4
\citep{blanton07}. For $z=1.4$
sources, the $g-$band ($\sim0.5\mu$m) redshifts into the 
$J-$band ($\sim 1.2\mu$m) and the $z-$band ($\sim0.9\mu$m) redshifts into the
$K-$band ($\sim 2.2\mu$m). As a result, fields with extensive
photometric coverage use observed-frame bandpasses that are very close to 
the rest-frame $g-$ and $z-$ bandpasses of interest, and so the $k-$corrections are
small. For our TBT diagnostic, we subtract the rest-frame $z-$band
from the rest-frame $g-$band to obtain the rest-frame $g-z$ color;
hereafter, $^{0.0}(g-z)$.  

To determine the error associated with the $^{0.0}(g-z)$ color
for each sample, we randomly alter the magnitudes by an amount
consistent with the photometric uncertainties and re-run the \citet{blanton07}
software. We then determine the $1~\sigma$ standard
deviation on $\Delta = ~^{0.0}(g-z)_{\rm original} - ~^{0.0}(g-z)_{\rm
  random}$. For our SDSS sample, $\sigma_{\Delta} = 0.21$. For our
GOODS-N/LH sample, $\sigma_{\Delta} = 0.07$. The high-quality
photometry and extensive coverage of our GOODS-N/LH fields ensures
these low $k-$correction errors. 

\citet{chilingarian10} compare the \citet{blanton07} kcorrect code
with their own to explore the systematic error associated with $k-$correction
software. For $k-$corrected optical colors (e.g., $^{0.0}[g-r]$, $^{0.0}[r-z]$), the 
differences between the resulting colors from the two codes are
relatively small ($\sim 0.05$ or less). Therefore, for our study,
systematic errors are not significant. However, we note here for
completeness that for $k-$corrected
colors in which one band is in the rest-frame IR (e.g., $^{0.0}[r-H]$,
$^{0.0}[r-K]$), the differences can be significant ($\sim 0.15$; see
their Fig.~7), and the systematic error need be taken into account.  


\section{A New $z<1.4$ Emission-Line Ratio Diagnostic}
\label{TBT}


\begin{figure}[!htp]
\epsscale{1.2}
\plotone{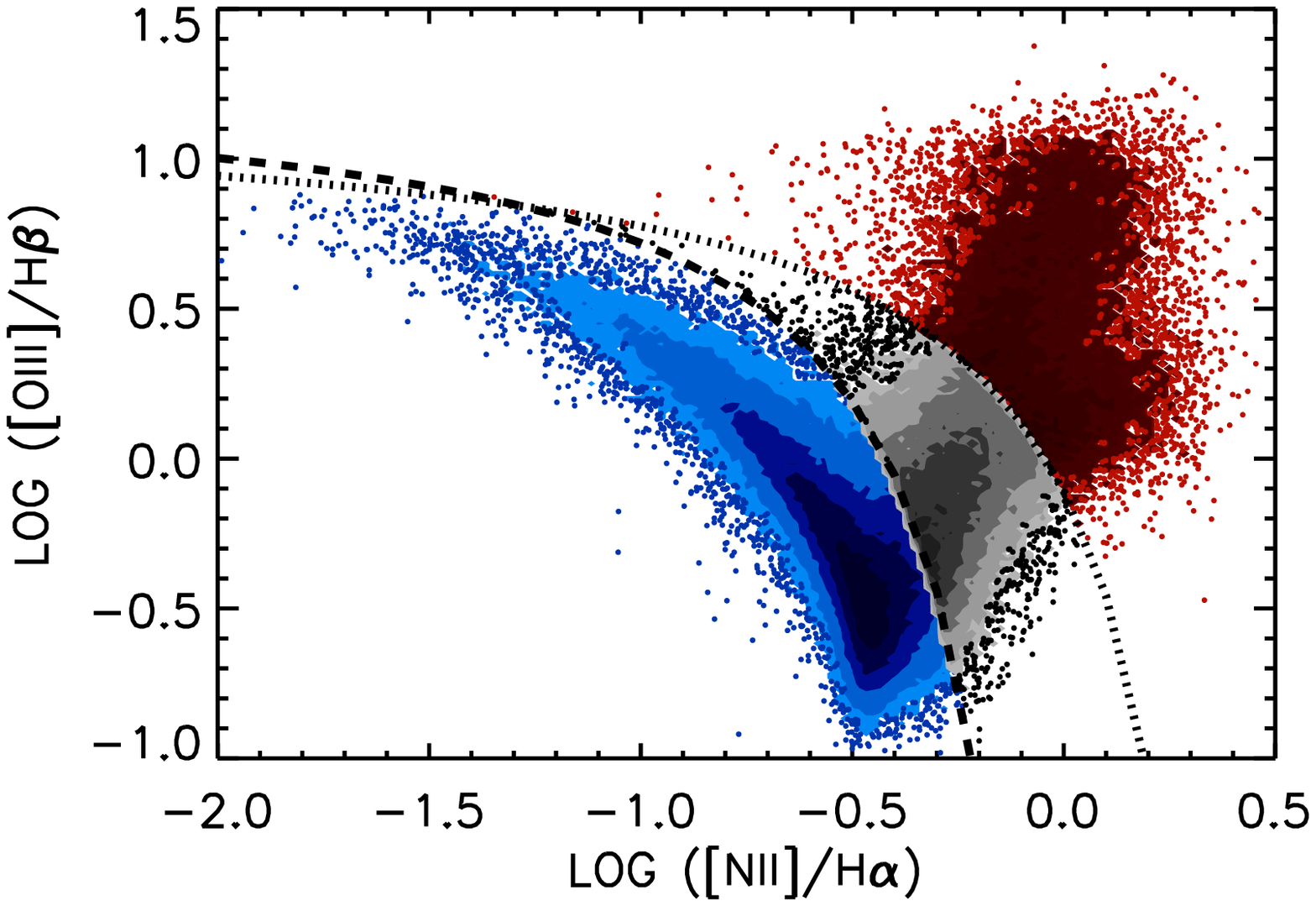}  
\caption{BPT diagram -- $\log$([NII]/H$\alpha$)
  vs.~$\log($[OIII]/H$\beta$) -- for the SDSS
  galaxies that have [OIII], H$\beta$, [NII], and
H$\alpha$ with SNR~$>5$. The dashed curve shows the
  \citet{kauffmann03} empirical division between star-forming galaxies and
  AGNs. The dotted curve shows the \citet{kewley01} theoretical
  division.  As discussed in the text, we represent the SDSS
data using a combination of a two-dimensional histogram and plotted
points. BPT-SF are shown in blue, BPT-comp in gray, and BPT-AGN in
red.}
\label{SDSS_BPT}
\end{figure}


\begin{figure}[!htp]
\epsscale{0.9}
\plotone{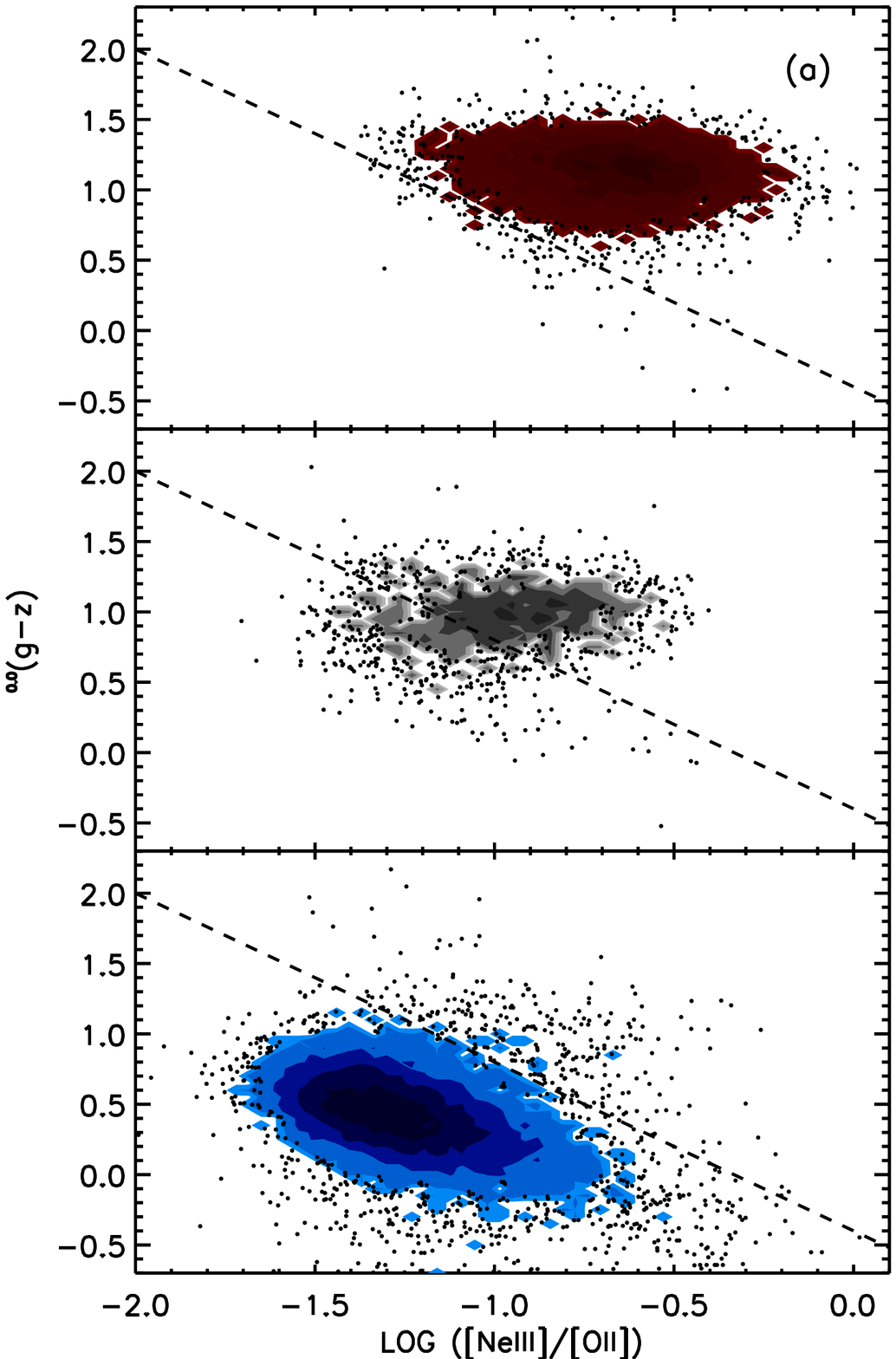}  
\plotone{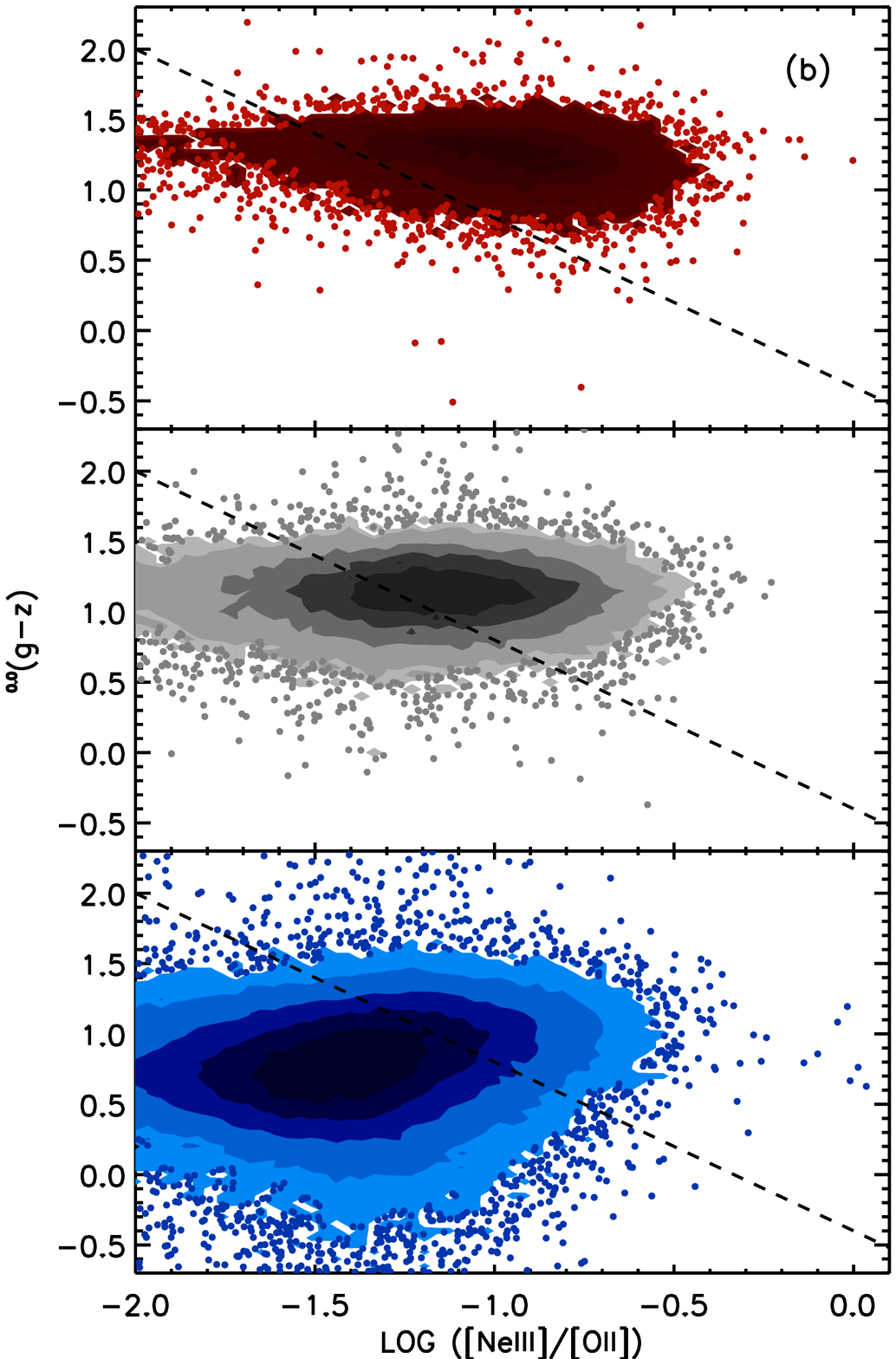}  
\caption{(a) TBT diagram -- \gminz\ color vs.~$\log$([NeIII]/[OII]) -- 
  for the SDSS galaxies that have [OIII], H$\beta$, [NII],
H$\alpha$, [NeIII], and [OII] fluxes with SNR~$>5$. The top, middle, and bottom
  panels show  the positions of the BPT-AGN, BPT-comp, and BPT-SF,
  respectively. The dashed 
  line provides the empirical separation maximizing the fraction of
  BPT-AGN to total population in the 
upper-right of the diagram (see Eq.~\ref{hizDiv}). (b) Same as (a) but
for SDSS galaxies that have  [OIII], H$\beta$, [NII],
H$\alpha$, and [OII] with SNR~$>5$, but [NeIII] has SNR~$<5$. For the
15\% with [NeIII]~$\le 0$, we set [NeIII] equal to the $1~\sigma$
error.}
\label{SDSS_TBT}
\end{figure}

Figure \ref{SDSS_BPT} shows the BPT diagram (based on flux ratios of
the specified lines) applied to our SDSS BPT sample. The dotted curve indicates the
\citet{kewley01} theoretical division and the dashed curve
indicates the \citet{kauffmann03} empirical 
division between AGNs and star-forming galaxies, as discussed in the
Introduction. In this and subsequent figures we represent the SDSS
data using a combination of a two-dimensional histogram and plotted
points. We histogram the data where more than four data points fall in
an individual pixel (the size of the pixel is determined by dividing
the plot into 150x150 bins) and plot it as individual points otherwise. The
histograms have been square root scaled for better visibility. Each
step in contour level represents 20\% fewer sources.  The
darkest contours correspond to 391 sources, 42 sources, and 151 sources
for the BPT-SF, BPT-AGN, and BPT-comp, respectively. 

We find that 69\% of our SDSS BPT sample lie below the
\citet{kauffmann03} division, in the BPT-SF
regime, 11\% lie above the \citet{kewley01} division, in
the BPT-AGN regime, and 20\% lie in between the two
divisions, in the BPT-comp regime. If instead we apply the BPT diagnostic to
our SDSS TBT sample (i.e., SDSS galaxies that have [OIII], H$\beta$, [NII],
H$\alpha$, [NeIII], and [OII] fluxes with SNR~$>5$), we find that 60\% are
BPT-SF, 32\% are BPT-AGN, and 8\% are BPT-comp. 

The BPT diagram can only be used to identify AGNs out to $z=0.5$, the
redshift at which [NII] leaves the optical spectral window. The ratio
of [NeIII] to [OII] is a good 
candidate for pushing optical narrow emission-line ratio diagnostics
to higher redshifts because both lines are relatively strong, lie in
the blue end of the spectrum (measurable in optical spectra out to $z<1.4$), and
are close in wavelength (thus avoiding reddening effect
issues). Furthermore, [NeIII] has a significantly higher ionization
potential (63.45 eV) than [OII] (35.12 eV). As a result, the
[NeIII]/[OII] ratio is higher in AGNs than in star-forming galaxies. We also note that
the ionization potential for [NeIII] is 
significantly higher than for [OIII], whose ionization potential is 54.94 eV. 
In Section \ref{disc} we
discuss how AGN selection using a [NeIII]-based diagnostic may be more
discerning than one based on [OIII] as a result of this higher
ionization potential (requiring a harder ionizing flux). 

In Figure \ref{SDSS_TBT}(a)
we plot our TBT diagnostic -- \gminz\ color versus
$\log$([NeIII]/[OII]) -- for our SDSS TBT sample, color-coded according
to the BPT classifications in Figure \ref{SDSS_BPT}. The top, middle,
and bottom panels show 
the locations of the BPT-AGN, BPT-comp, and BPT-SF, respectively. The
darkest contours correspond to 71 sources, 16 sources, and 68 sources
for the BPT-AGN, BPT-comp, and BPT-SF, respectively. The
BPT-SF show a trend with color in that the bluer BPT-SF exhibit higher
values of [NeIII]/[OII]. This provides a separation in color-space between the
BPT-AGN and the BPT-SF with high values of [NeIII]/[OII]. This trend with color is
likely due to the fact that bluer galaxies are more metal poor (see
Fig.~7 in \citealt{tremonti04}) and hence have harder stellar
radiation fields (higher [NeIII]/[OII]; \citealt{shi07}). On the other hand,
few BPT-AGN reside in very 
blue galaxies. As discussed in detail in Section 3.3 of \citet{yan11}, nearly all
BPT-AGN are found in red galaxies or in galaxies with intermediate colors
between red and blue. 

The dashed line designates
\begin{equation} 
\label{hizDiv}
^{0.0}(g-z)=-1.2\times\log(\rm{[NeIII]/[OII]})-0.4 \,, 
\end{equation}
our empirical separation maximizing the fraction of BPT-AGN to total
population in the upper-right of the diagram. Hereafter, we refer
to the sources in the upper-right (lower-left) of our TBT diagnostic as
TBT-AGN (TBT-SF). 

We find that 98.7\%
of the BPT-AGN lie in the TBT-AGN regime and 97.2\% of the BPT-SF
lie in the TBT-SF regime. Likewise, we find that only 3.5\% of the sources in the
TBT-AGN regime are BPT-SF and 1\% 
of the sources in the TBT-SF regime are BPT-AGN.  The BPT-comp lie on either
side of the division, with 68.6\% in the TBT-AGN regime. Overall the
BPT-comp constitute 5\% and 16\% of the TBT-SF and TBT-AGN,
respectively. 

There are 200,712 galaxies in our SDSS BPT sample that have [OII] with
SNR~$>5$ but [NeIII] with SNR~$<5$. In Figure \ref{SDSS_TBT}(b) we
show the TBT diagram for these sources. For the 15\% with [NeIII]~$\le
0$, we set [NeIII] equal to the $1~\sigma$ error. The
darkest contours correspond to 94 sources, 170 sources, and 546 sources
for the BPT-AGN, BPT-comp, and BPT-SF, respectively. We find that $\sim85$\% of
the BPT-SF lie within the TBT-SF regime, $\sim86$\% of the BPT-AGN
lie within the TBT-AGN regime, and $\sim62$\% of the BPT-comp lie
within the TBT-AGN regime. The
trends follow those for our SDSS TBT sample. The 8\% of SDSS BPT
galaxies for which neither [OII] nor [NeIII] has 
SNR~$>5$ and the $<1$\% of SDSS BPT galaxies for which [NeIII] has
SNR~$>5$ but [OII] has SNR~$<5$ are not considered here. 

We have trained our TBT diagnostic on the SDSS galaxies, which have
$\langle z \rangle \sim 0.1$. Because our TBT diagnostic can be used
with optical spectra out to $z=1.4$, we examine the impact of
metallicity evolution with redshift on our empirical separation 
between TBT-AGN and TBT-SF. Galaxy metallicities decrease by a factor of $\sim
0.3$ dex between the local value and the value at $z\sim 2$
(\citealt{erb06}; see also \citealt{cowie08,kewley08,zahid11}). For an
M$_{\rm star} = 10^{10}$M$_\sun$ galaxy, this corresponds to a shift
from $12+\log$(O/H) $=8.6$ to 8.3 (note that more massive galaxies
undergo less metallicity evolution). Using
the \citet{shi07} relation between metallicity
and $\log$([NeIII]/[OII]), we find that, in this case,
$\log$([NeIII]/[OII]) shifts by only
$\sim20$\%, from $-0.89$ to $-0.72$. Applying this 20\% increase in
the ratio of [NeIII] to [OII] to all SDSS BPT-SF, we find that an additional 
5\% move to the TBT-AGN regime. This corresponds to an increase of
only 5.5\% in the number of TBT-AGN that are BPT-SF. Similarly, there
is only a 1.8\% increase in the number of TBT-AGN that are
BPT-comp. Because the impact 
is relatively small, in this article we do not consider any
metallicity evolution with redshift in our empirical separation
between TBT-SF and TBT-AGN. 

We also considered the impact of color evolution on our TBT
diagnostic. At higher redshifts, galaxies are bluer as a result of
higher specific star formation rates. Purely passive evolution models
\citep{bruzual03} with an
instantaneous burst and a \citet{chabrier03} initial mass function
predict a $\Delta ^{0.0}(g-z) \sim 0.24$ between $z=0$ and 1.4, for a
formation redshift of 5. Applying this color evolution to all SDSS BPT-SF,
the impact is in our 
favor. The BPT-SF move down the y-axis in our TBT diagnostic to lie even
further below our empirical separation, in the 
TBT-SF regime.  Applying this color evolution to all
SDSS BPT-AGN, we find that an additional 3.6\% move to the TBT-SF
regime. This corresponds to an increase of 2.7\% in the number of TBT-SF
that are BPT-AGN. Similarly, there is only a 2.9\% increase in the
number of TBT-SF that are BPT-comp. Because the impact is small, in
this article we do not consider any color evolution with redshift in
our empirical separation between TBT-SF and TBT-AGN. 



\section{Comparing the TBT Diagnostic with an X-ray Selection of AGNs}
\label{xray}


\begin{figure}[!htp]
\epsscale{1.2}
\plotone{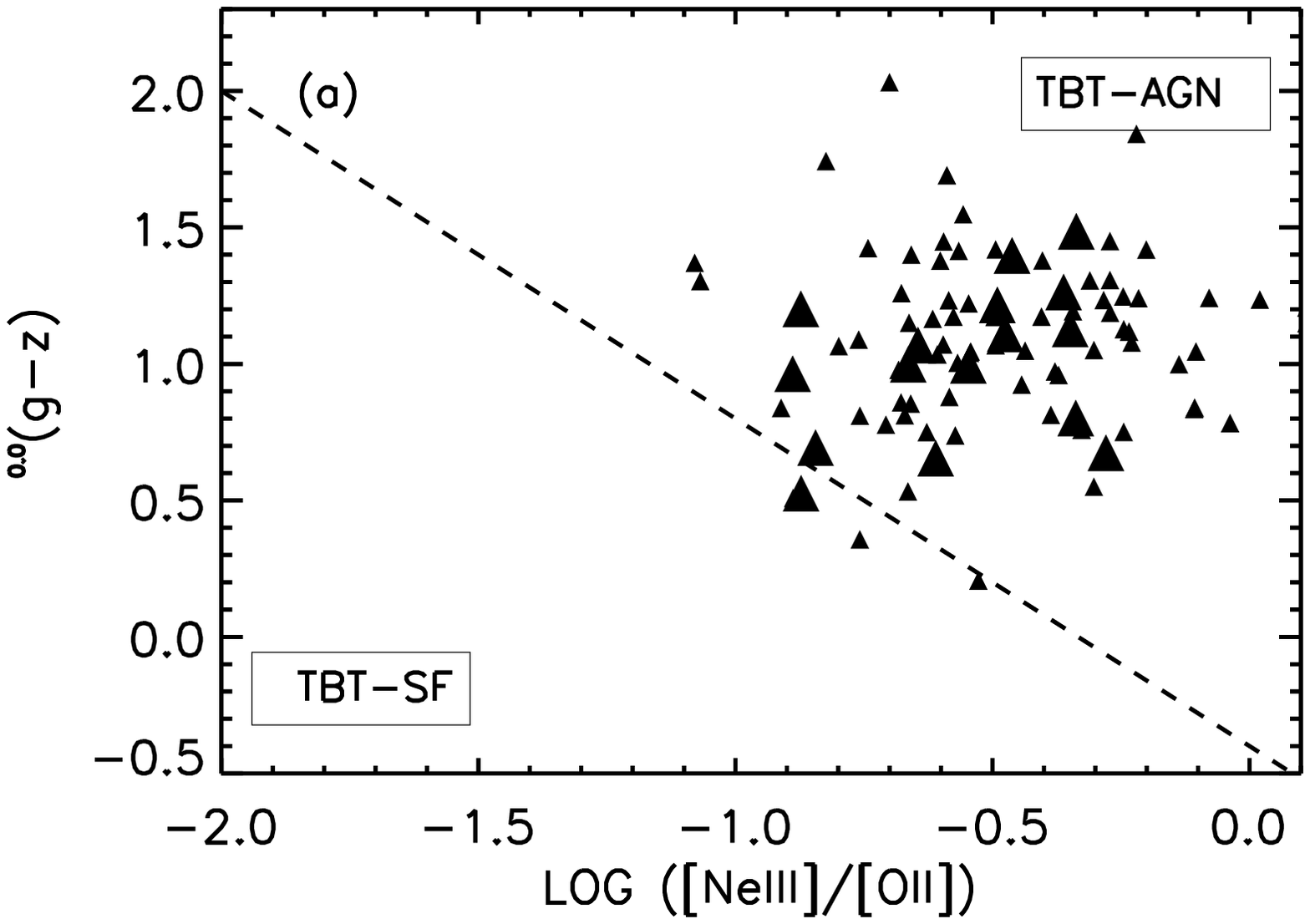} 
\plotone{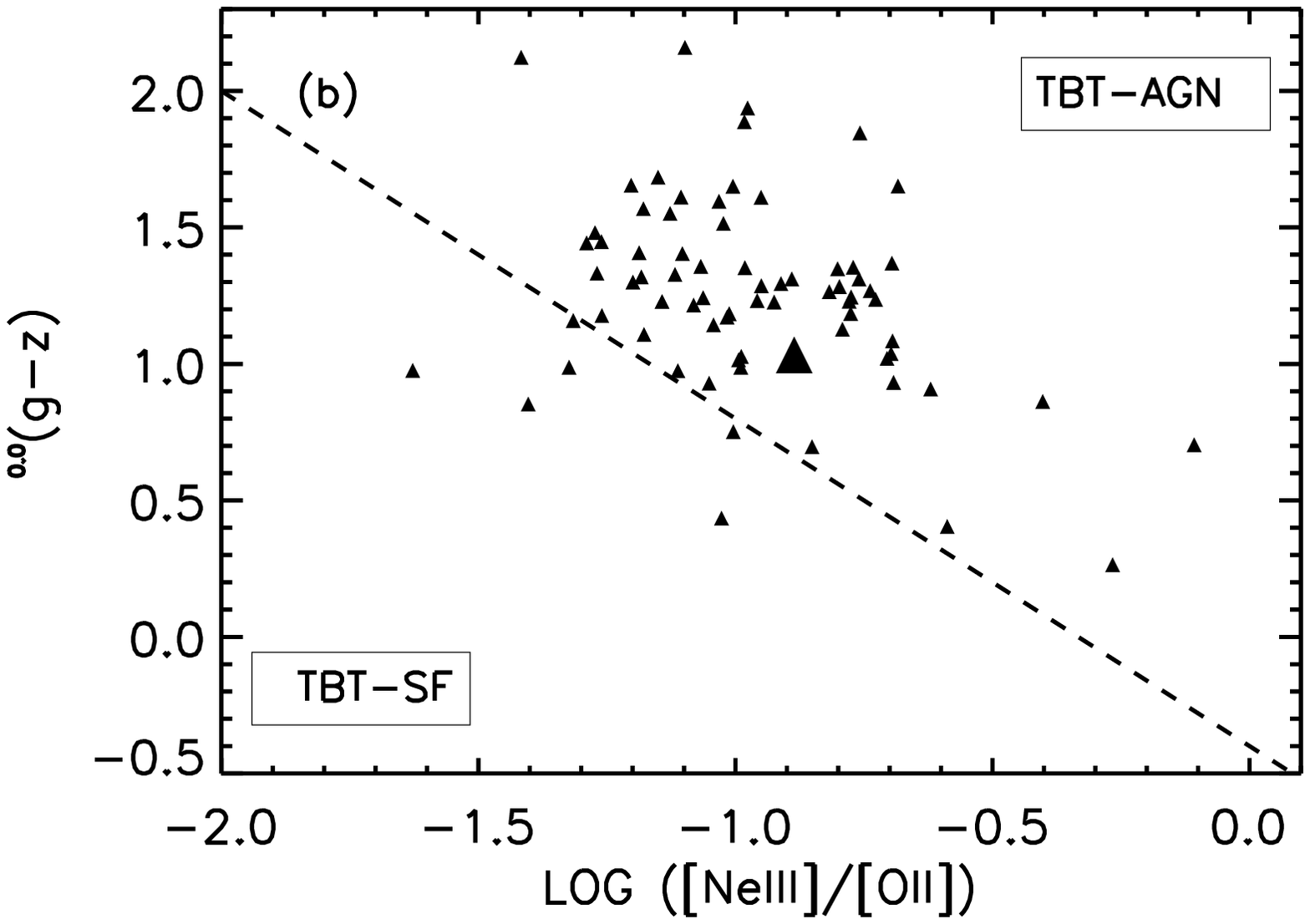} 
\caption{(a) TBT diagram -- \gminz\ color vs.~$\log($[NeIII]/[OII]$)$
  -- 
  for the X-ray selected non-BLAGNs in our OPTX sample with
  $0.3<z<1.4$ for which the [NeIII] and [OII] fluxes have a SNR~$>5$. \emph{Small
    black triangles}---non-BLAGNs
  with $10^{42}<L_X<10^{44}$~\ergss. \emph{Large black triangles}---non-BLAGNs
  with $L_X>10^{44}$~\ergss. The dashed line provides the empirical
  separation maximizing the fraction of BPT-AGN to total population in
  the upper-right of the diagram (see Figure \ref{SDSS_TBT} and
  Eq.~\ref{hizDiv}). (b) Same as (a) but for the OPTX X-ray selected non-BLAGNS for
  which [OII] has a SNR~$>5$ but [NeIII] has a SNR~$<5$. For the
6\% with [NeIII]~$\le 0$, we set [NeIII] equal to the $1~\sigma$
error.}
\label{OPTX_TBT}
\end{figure}

In \citet{trouille10} we found that only a little over half (52\%) of the X-ray selected
non-BLAGNs in our OPTX sample lie in the BPT-AGN regime of the BPT diagram. These
sources form a sequence 
similar to that of the BPT-AGN, emerging from the HII region
sequence and extending to the upper-right hand side of the BPT
diagram. 

Of the X-ray selected non-BLAGNs in our OPTX sample, 20\% are
misidentified as BPT-SF, i.e., as pure star-forming galaxies
(\citealt{trouille10}; see also
\citealt{winter10} for evidence of this in the \emph{Swift} BAT
sample). Increased extinction does not account for these. Instead, in
\citet{trouille10} we note that these misidentified sources have lower 
$L_{\rm [OIII]}/L_X$ ratios than those that lie in the
BPT-AGN regime. We postulate that the low
forbidden emission line strengths in the misidentified sources are a
result of the complexity of the structure of the narrow-line region,
which causes ionizing photons from the central engine to not be
absorbed. 

The misidentification of X-ray selected AGNs as star-forming galaxies is a
potential issue for all optical emission-line diagnostic diagrams,
including our TBT diagram. In Figure \ref{OPTX_TBT}(a) we plot
the TBT diagram -- \gminz\ color versus
$\log$([NeIII]/[OII]) -- for the $0.3<z<1.4$ X-ray selected
non-BLAGNs in our OPTX sample that have [NeIII] and [OII] with
SNR~$>5$. We find that $97$\% (100/103) of 
our X-ray selected AGNs lie in the TBT-AGN regime. Thus, the TBT diagnostic does
a much better job of correctly identifying X-ray selected AGNs than
the BPT diagnostic, misidentifying only $3$\% as TBT-SF compared to the
BPT diagnostic's misidentification of 20\% as BPT-SF. 

As mentioned in Section \ref{optx}, there are 94 $0.3<z<1.4$ OPTX
non-BLAGNs that do not fulfill the criteria of having both [NeIII] and
[OII] with SNR~$>5$. Sixteen of these have neither [OII] nor [NeIII]
with a SNR~$>5$. In Figure \ref{OPTX_TBT}(b) we show the remaining 78
OPTX X-ray selected non-BLAGNs for which [OII] has a SNR~$>5$ but
[NeIII] has a SNR~$<5$. No sources
have only [NeIII] with a SNR~$>5$.  We find that 92\% (72/78) lie in
the TBT-AGN regime. This supports our results for the OPTX sources in
which both [NeIII] and [OII] have a SNR~$>5$.  


\section{Verification of the TBT Diagnostic: Stacking Analyses}
\label{stack}

\begin{figure}[!htp]
\epsscale{1.2}
\plotone{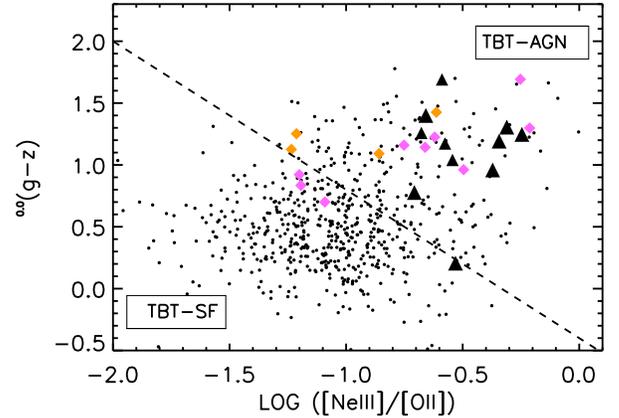} 
 \caption{TBT diagram -- \gminz\ color vs.~$\log$([NeIII]/[OII]) -- for the GOODS-N
  $0.3<z<1.4$ sources for which the
  [NeIII] and [OII] fluxes have a SNR~$>5$. \emph{Large black
    triangles}---non-BLAGNs cross-listed in the \emph{Chandra} catalog with
  $L_X>10^{42}$~\ergss. \emph{Orange diamonds}---non-BLAGNs
  cross-listed in the \emph{Chandra} catalog, but with
  $L_X<10^{42}$~\ergss. These sources are significantly detected in both the
  $0.5-2~\rm keV$ band and $2-8~\rm
  keV$ bands. \emph{Magenta diamonds}---non-BLAGNs cross-listed
  in the \emph{Chandra} catalog, but with $L_X<10^{42}$~\ergss. These sources
  are significantly detected
  in the $0.5-2~\rm keV$ band but not the $2-8~\rm keV$ band.
  \emph{Black circles}---remaining sources. The dashed line provides the empirical
  separation maximizing the fraction of BPT-AGN to total population in
  the upper-right of the diagram (see Figure \ref{SDSS_TBT} and Eq.~\ref{hizDiv}).}
\label{goods_NeIII}
\end{figure}

The GOODS-N sample of spectroscopically
observed galaxies with deep \emph{Chandra} and \emph{Spitzer} imaging
out to $z=1.4$ provides an ideal pilot for confirming the dominance of
AGN activity in the
TBT-AGN and star formation activity in the TBT-SF. 

Figure \ref{goods_NeIII} shows our TBT diagnostic for the 642 $0.3<z
\le 1.4$ GOODS-N sources for which the
  [NeIII] and [OII] fluxes have a SNR~$>5$. We have excluded sources with
  broadened emission lines (FWHM~$>2000$~km~s$^{-1}$) from this
  analysis. The dashed line 
indicates our empirical division between the TBT-AGN and TBT-SF (see
Figure \ref{SDSS_TBT} and Eq.~\ref{hizDiv}). We find that 189
GOODS-N sources lie in the TBT-AGN regime,
corresponding to a space density of $4.6\times10^{-5}$~Mpc$^{-3}$
optically selected GOODS-N TBT-AGN. 

There are 11 GOODS-N sources in our TBT diagram that are cross-listed
as non-BLAGNs 
in the CDFN catalog with $L_X>10^{42}$~\ergss~(black triangles). As
discussed in Section \ref{xray}, our TBT diagnostic has a much lower rate
of misidentifying X-ray selected AGNs as star-forming galaxies than the BPT
diagnostic ($\sim3$\% versus 20\%). We use our OPTX X-ray selected
sample of AGNs for that analysis. Since the GOODS-N field lies within
the CDFN, one of our OPTX fields, it is no surprise that only one of
the $L_X>10^{42}$~\ergss~non-BLAGNs (black triangles) in Figure
\ref{goods_NeIII} is misidentified as a TBT-SF. 

In the following section we perform X-ray and IR stacking analyses in order to
investigate whether the remaining $L_X<10^{42}$~\ergss~TBT-AGN harbor
obscured AGN activity, verifying the validity of our TBT-AGN selection. We
also perform stacking analyses of our TBT-SF in order to verify the
dominance of star formation activity in these sources. The effective X-ray
spectral slope ($\Gamma_{\rm eff}$; derived from the ratio of the
$2-8~\rm keV$ band to the $0.5-2~\rm keV$ band) is indicative of different source
types. Unobscured AGNs exhibit $\Gamma_{\rm eff} \ge 1.7$. In obscured
AGNs, photoelectric
absorption of X-ray soft photons by neutral gas along the line of
sight causes a flattening of the slope, such that $\Gamma_{\rm eff} <
1.7$ (although see \citealt{trouille09} for a discussion of
discrepancies between X-ray and optical spectral type). At the same
time, X-ray binaries exhibit a range in X-ray spectral slopes. LMXBs, which are
associated with old stellar populations, tend to be softer in
X-rays, with $\Gamma>1.7$.  HMXBs, which are associated with ongoing star
formation, tend to be harder in X-rays, with $\Gamma
= 0.5-1$ \citep{colbert04}. See \citet{fabbiano06} and
\citet{remillard06} for detailed reviews of X-ray binary
populations. In order to distinguish between these scenarios, we
perform an IR stacking analysis in Section \ref{IRstack}.  

We restrict our
analysis to lower redshift sources ($z<0.7$) in order to not be misled
by the automatic softening of the effective X-ray 
spectral slope as one observes to higher redshifts (i.e., as the
redshift increases, the $0.5-2~\rm keV$ and $2-8~\rm keV$ bands are
sampling higher energies whose photons can more easily penetrate
obscuring material). 

\subsection{X-ray Stacking}
\label{GOODSxray}

Quantitatively, if stacking a sample of `$n$' objects yields `$N_s$'
counts in a signal aperture of area `$A_s$' and `$N_b$' smoothed
background counts in that
same area `$A_s$', then the mean number of source counts per object
in the signal region is
\begin{equation} 
\langle N \rangle _{src} = \frac{1}{n} [N_s - N_b] \,.
\end{equation}

We obtain an estimate of the mean spectral slope of the detected
signal by performing the stacking in two energy bands, soft ($0.5-2
~\rm keV$) and hard ($2-8~\rm keV$), and deriving an effective
power-law photon index, $\langle \Gamma_{\rm eff} \rangle$. 

Here we use the STACKFAST X-ray stacking program
\citep{hickox07}.  In STACKFAST, `$A_s$' is defined as
the area enclosed within $r_{90}$ from the source 
position, where $r_{90}$ is an approximation of the 90\% point-spread
function (PSF) energy encircled radius at 1.5 keV, and varies
as\footnote[1]{http://cxc.harvard.edu/proposer/POG}:
\begin{equation}
r_{90} = 1\arcsec + 10\arcsec (\theta / 10 \arcmin)^2 \,,
\end{equation}
with $\theta$ equal to the off-axis angle. In order to maximize the
number of source counts, rather than limiting the stacking to only the
central $6\arcmin$ as in \citet{hickox07}, we use the central
$10\arcmin$ around the pointing center for each observation. 

A few bright sources would dominate our
estimate of the mean spectral shape, so we exclude
from our analysis sources that lie close to or are associated with a known
X-ray detected source, hereafter called coincidental contaminants \citep[see
also][]{hickox07,georgantopoulos08,fiore08}. To this end, we first
applied a mask to all known X-ray detected source positions. We
used $3\times r_{90}$ (see Eq.~3) as the mask aperture
radius.

We cross reference the `$A_s$' for our sources of interest with the
X-ray photon locations from each \emph{Chandra} pointing (i.e., each
OBSID event file). 
We then create smoothed $0.5-2~\rm keV$ and $2-8~\rm keV$ background maps
using the CIAO WAVDETECT task and
determine the background counts within the same `$A_s$'.  Subtracting
this background and dividing by the number of sources being stacked,
we obtain the average X-ray signal in counts per source. We create
exposure maps using custom routines (A.~Vikhlinin, private communication)
and determine the total exposure time for each source
being stacked. Dividing the total counts in the stacked source by the
total exposure time, we derive the average count rate (counts per second) for the
stacked source. We use the ratio of the $2-8~\rm keV$ count rate to the
$0.5-2~\rm keV$ count rate to derive $\langle \Gamma_{\rm eff} \rangle$. 

Count uncertainties are calculated using the approximation $\sqrt{X +
0.75} + 1$, where $X$ is the number of counts in a given band
\citep{gehrels86}. We set a significance threshold of $3~\sigma$. 
Uncertainties in the hardness ratio and
$\langle \Gamma_{\rm eff} \rangle$ are derived by propagating these count rate errors. 

Table \ref{table_stackTBT} shows the results from our stacking
analysis for the GOODS-N TBT-SF and TBT-AGN. Column 2 lists the total
number of sources in 
each of these categories. Column 3 provides the total number of sources
used in the stacking analysis, after excluding individually X-ray detected
sources and coincidental contaminants. Columns 4 and 5 state
fluxes and detection significance for 
the given X-ray band. We consider that stacked sources with
$<3~\sigma$ detection are not significantly detected. Column 6 provides the
$\langle \Gamma_{\rm eff} \rangle$ value for the stacked source. In
column 7 we list the volume-weighted redshift for the stacked source
and in column 8 we list the derived $2-8~\rm keV$ luminosity based on
the $f_{2-8~\rm keV}$ and the volume-weighted redshift.  

We provide specifics for each category in the following subsections. 


\begin{table*}[htp]
\small
\caption{TBT X-ray Stacking Analysis Results \label{table_stackTBT}}
\begin{tabular}{lccccccc}
\tableline\tableline
Category   & Total \#     &  \# used in stack    &  $f_{0.5-2~\rm
  keV}$\tablenotemark{a}  & $f_{2-8~\rm keV}$\tablenotemark{a} & $\langle 
\Gamma_{\rm eff} \rangle$ & z\tablenotemark{b} & $\log L_{2-8~\rm keV}$
[erg~s$^{-1}$] \\
(1)             & (2)             & (3)                     & (4) & (5) & (6) & (7) & (8) \\
\tableline
TBT-SF\tablenotemark{c}     & 155  & 148  & $0.62 \pm 0.13$
($4.7~\sigma$) & $2.50 \pm 1.1$ ($2.2~\sigma$) &  $1.5_{-0.4}^{+0.7}$\tablenotemark{d} & 0.58 & $40.53$\tablenotemark{d} \\
TBT-AGN\tablenotemark{c} & 72  & 54 & $1.13 \pm 0.2$ ($5.4~\sigma$) & $5.87 \pm 1.4$
($4~\sigma$) &  $1.0_{-0.3}^{+0.3}$ & 0.58 & 40.86 \\
\tableline\tableline
\end{tabular}
\tiny\\
$^a$In units of $10^{-17}$~erg~cm$^{-2}$~s$^{-1}$.\\
\vspace{-0.05in}$^b$Volume-weighted redshift for the stacked source.\\
\vspace{-0.05in}$^c$Only including sources with $z<0.7$.\\
\vspace{-0.05in}$^d$We caution that this is based on a low significance $2-8~\rm keV$ signal.
\normalsize
\end{table*}

\subsubsection{TBT-SF}
\label{goods_tbtsf}

There are 155 $z<0.7$ GOODS-N TBT-SF. As discussed above, we restrict our
analysis to these lower  redshifts in order to not be misled by the automatic softening
of the effective X-ray 
spectral slope as one observes to higher redshift. Only one of these
$z<0.7$ GOODS-N TBT-SF is directly X-ray 
detected, and then only in the $0.5-2~\rm keV$ band. Using
the $2-8~\rm 
keV$ flux limit for the CDFN image (see Table \ref{table_fluxlim}), we
determine an upper limit to its hardness ratio and find that it is
X-ray soft, with $\Gamma > 1.7$. 

After excluding this known X-ray source, as well as coincidental contaminants,
we use the STACKFAST program to determine the average X-ray signal in
the remaining 148 TBT-SF. The stacked source is significantly detected
in the $0.5-2~\rm keV$ band ($4.7~\sigma$) but not in the $2-8~\rm keV$
band ($2.2~\sigma$). We find a $\langle \Gamma_{\rm eff}
\rangle =1.5_{-0.4}^{+0.7}$, although we caution that this is based on
a low-significance $2-8~\rm keV$ signal. Given the large
uncertainties, the stacked source could be X-ray soft ($\Gamma > 1.7$). 

\subsubsection{TBT-AGN}
\label{goods_tbtagn} 

There are 72 $z<0.7$ GOODS-N TBT-AGN. Six are directly X-ray
detected in both the $0.5-2~\rm keV$ and $2-8~\rm keV$ bands. Three of
these are obvious X-ray selected AGNs with 
$L_X>10^{42}$~erg~s$^{-1}$. The 
three remaining X-ray detected sources have
$L_X<10^{42}$~erg~s$^{-1}$. All three are X-ray hard, with $\Gamma < 1.4$.

After excluding these known X-ray sources, as well as coincidental
contaminants, 
we use the STACKFAST program to determine the average X-ray signal in
the remaining 54 TBT-AGN. The stacked source is significantly detected
in both bands ($5.4~\sigma$ in the $0.5-2~\rm keV$ band and
$4.0~\sigma$ in the $2-8~\rm keV$ band). We find a $\langle
\Gamma_{\rm eff} \rangle =1.0_{-0.3}^{+0.3}$, consistent with being
X-ray hard ($\Gamma < 1.4$). 

\subsubsection{Monte-Carlo Simulation}
\label{montecarlo}

We carried out a series of Monte Carlo (MC) stacking simulations to assess
false-detection probabilities empirically. For each category we performed 1000
trials and used the same number of stacked sources and the same procedure
as in the original stacking, albeit with random RA and Dec positions. 

In Section \ref{goods_tbtsf} we found that the TBT-SF are
significantly detected in the $0.5-2~\rm keV$ band but not in the
$2-8~\rm keV$ band. In Section \ref{goods_tbtagn} we found that the 
TBT-AGN are significantly detected in both the $0.5-2~\rm keV$ and
$2-8~\rm keV$ bands. 

Our MC simulations yield a 0.1\% probability of generating the observed
$0.5-2~\rm keV$ flux for our stacked TBT-SF source and a 52\%
probability of generating the observed $2-8~\rm keV$ flux. Our MC
analysis confirms that the TBT-SF clearly have an excess of
$0.5-2~\rm keV$  counts, well above those we obtain randomly. The 
$2-8~\rm keV$ signal is within the noise.  Thus, the observed X-ray
softness in our TBT-SF appears to be a reliable result. 

Our MC simulations for
the TBT-AGN yield a 0\% probability that we would
see the observed $0.5-2~\rm keV$ flux or the observed $2-8~\rm
keV$ flux. Our MC analysis confirms we are recovering a real signal in
both bands.  None of our MC simulations for the TBT-AGN
result in both the $0.5-2~\rm keV$ and $2-8~\rm keV$ bands detected
at a $3~\sigma$ level or greater. In other words, there is a $\sim 0$\%
false-alarm rate for our stacked TBT-AGN signal. Thus, the
observed X-ray hardness in our TBT-AGN appears to be a reliable result. 

\subsection{IR Stacking}
\label{IRstack}


\begin{figure}
\epsscale{1.2}
\plotone{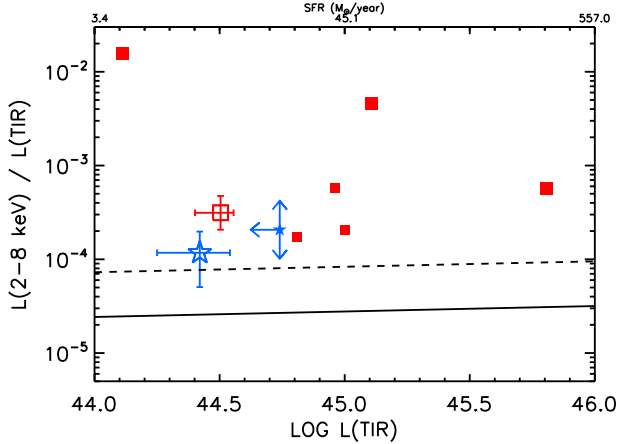} 
\caption{$L_{2-8~\rm keV}/L_{TIR}$ versus $\log L_{TIR}$ for our
  $0.3<z<0.7$ GOODS-N galaxies that have [NeIII] and [OII] fluxes with
  SNR~$>5$. \emph{Red filled squares} -- directly X-ray
 detected TBT-AGN. Large (small) symbols indicate sources with
 $L_X>10^{42}$~erg~s$^{-1}$ ($L_X<10^{42}$~erg~s$^{-1}$). \emph{Blue
   filled star} -- directly X-ray detected TBT-SF. The downward
 pointing arrow indicates that the source is undetected in the
  $2-8~\rm keV$ band and is assigned the $2-8~\rm keV$ flux limit (see Table 
  \ref{table_fluxlim}). The upward and leftward pointing 
  arrows indicate that the source is undetected in the \emph{Spitzer} $24\mu$m
  image and is assigned the 24$\mu$m flux limit (see Table
  \ref{table_fluxlim}). \emph{Red open square} -- stacked
  TBT-AGN, excluding directly X-ray detected sources. \emph{Blue open
    star} -- stacked TBT-SF, excluding directly X-ray detected
  sources. \emph{Solid
    line} -- expected ratio for the HMXB population in 
  a galaxy with the given SFR (see top axis),
  according to \citet{persic04}; \emph{dashed line} -- expected ratio
  for the overall star formation occurring in a galaxy with the given
  SFR, according to \citet{mineo11}.}
\label{LxIR}
\end{figure}

\begin{table}[htp]
\caption{Flux Limits \label{table_fluxlim}}
\begin{tabular}{lc}
\tableline\tableline
\emph{Chandra} $2-8~\rm keV$ Image  & Flux Limit (~erg~cm$^{-2}$~s$^{-1}$)   \\
\tableline
CDFN & $1.5\times10^{-16}$ \\
CLANS &  $3.5\times10^{-15}$\\
CLASXS &  $3.5\times10^{-15}$\\
\tableline
\emph{Spitzer} $24\mu$m Image & Flux Limit ($\mu$Jy)\\
\tableline
GOODS-N & 75\\
LH\tablenotemark{a} & 150 \\
\tableline\tableline
\end{tabular}\\
$^a$Includes both the CLANS and CLASXS fields.
\normalsize
\end{table}

In Section \ref{GOODSxray} we derived the average X-ray properties for the
GOODS-N TBT-SF and TBT-AGN using an X-ray stacking analysis. The
stacked TBT-AGN source was significantly 
detected in both the $0.5-2~\rm keV$ and $2-8~\rm keV$ bands and was
quite X-ray hard, with $\langle \Gamma_{\rm eff} \rangle =
1.0_{-0.3}^{+0.3}$.  This X-ray hard signal could be due either to AGN
activity with obscuration or HMXBs
associated with ongoing star formation.

Numerous studies have found that in star-forming galaxies without an AGN, the
total X-ray luminosity correlates with the SFR
\citep{nandra02,bauer02,ranalli03,grimm03,colbert04,persic04,hornschemeier05,persic07,rovilos09}. 
Furthermore, \citet{david92} found a linear relation between the far-infrared
(FIR) luminosity and the $0.5-4.5~\rm keV$ luminosities for a sample of
starburst galaxies observed by \emph{Einstein} \citep[see
also][]{fabbiano02}. \citet{ranalli03}
extended this study to the $2-10~\rm keV$ band using the \emph{ASCA}
and \emph{BeppoSAX}
satellites. \citet{persic04} determined the relation between the X-ray
and FIR luminosities for both the full contribution from
star formation activity and for HMXBs alone. A number of groups have
shown that AGNs and AGN-starburst composite galaxies lie above these
relations \citep{ptak03,alexander05,teng05,georgakakis07}. The
additional X-ray luminosity is attributed to AGN activity in these
galaxies.  

Following these previous studies, here we use the IR properties of our
TBT-AGN to determine 
whether the X-ray hard signal is due to AGN activity with obscuraiton
or HMXBs. We also check whether the X-ray soft signal in our TBT-SF 
is consistent with pure star-forming galaxies. In \citet{trouille09} we
determined the \emph{Spitzer} $24 \mu$m fluxes and
luminosities for our OPTX X-ray selected AGNs. We follow the same
procedure here to determine $f_{24\mu\rm m}$ and $L_{24\mu\rm m}$ for our
GOODS-N galaxies. 

To determine $f_{24\mu\rm m}$ for our stacked TBT-AGN and stacked
TBT-SF, we use the publicly available IAS Stacking Library IDL
software \citep{bethermin10}. The IAS Stacking software uses a
DAOPHOT-type photometry IDL procedure, APER, with a preset PSF for
the \emph{Spitzer} $24\mu$m band ($13 \arcsec$ for the object aperture
and sky radii of $20\arcsec - 32\arcsec$). A median stacking is
preferable to mean stacking because the 
median analysis is more stable and robust to small numbers of bright
sources. Using the volume-weighted redshift for our stacked source, we
transform $f_{24\mu\rm m}$ into $L_{24\mu\rm m}$. We then use the
\citet{rieke09} eq.~A6 to transform $L_{24\mu\rm m}$ into the total
infrared luminosity, $L_{\rm TIR}$.  

The IAS Stacking software provides the $1~\sigma$ standard deviation
on the stacked flux. However, given that the \emph{Spitzer} image
resolution is low, we need to consider the likelihood of
misidentifications and overlap. To assess how well this $1~\sigma$
standard deviation reflects the contamination from spurious signals,
we carry out a series of Monte Carlo stacking. We perform 1000
trials and use the same number of stacked sources and follow the
same procedure as in the original stacking. The only difference is
that we use random RA and Dec positions. For both our stacked
TBT-AGN and our stacked TBT-SF, $>99$\% of our simulations result
in stacked fluxes less than our $1~\sigma$ error. While $<10$\% of
the random RA, Dec positions in each simulation do overlap with real
$24\mu$m sources (as expected given the low image resolution),
because we use a median stacking, 
these spuriously high fluxes are excluded from the stacked signal. 

In Figure \ref{LxIR} we plot the ratio of the
X-ray to total IR luminosities, $L_X/L_{\rm TIR}$, versus $L_{\rm TIR}$
for our TBT categories. The top axis shows the associated
SFR for a given $L_{\rm TIR}$, following \citet{rieke09}. The solid line shows the
expected ratio derived by \citet{persic04} for the HMXB population in a
galaxy with the given SFR. The dashed line shows the \citet{mineo11}
expected ratio for all star formation activity in a galaxy with the
given SFR. The
stacked TBT-SF (blue open star) was significantly detected in the
$0.5-2~\rm keV$ band ($4.7~\sigma$) but not in the $2-8~\rm keV$ band
($2.2~\sigma$). Nonetheless, we use the derived $L_X$ from
the stacking analysis (see Table
\ref{table_stackTBT}) and determine the $1~\sigma$ error on
$L_X/L_{\rm TIR}$ by propogating the errors on both the stacked X-ray
and IR signal.  The stacked TBT-SF $L_X/L_{\rm TIR}$ is
consistent with that expected for pure star-forming galaxies. This
corroborates what we find for the one individually X-ray detected
TBT-SF (blue filled star). Since this source is 
undetected in both the \emph{Spitzer} 24$\mu$m image and the
CDFN $2-8~\rm keV$ image, we assign it the flux limits for
these images (see Table \ref{table_fluxlim}) and use the arrows to
designate it as corresponding to upper limits. Given these
uncertainties, its $L_X/L_{\rm TIR}$ is not inconsistent 
with pure star-forming galaxies. 

The three X-ray detected TBT-AGN with
$L_X>10^{42}$~erg~s$^{-1}$ (large red filled squares) and the three
X-ray detected TBT-AGN with $L_X<10^{42}$~erg~s$^{-1}$
(small red filled squares) lie clearly 
above the expected range  for $L_X/L_{\rm TIR}$ for pure star-forming
galaxies. The stacked TBT-AGN (red open square) also lies well above  
this range. We determine the $1~\sigma$ error on $L_X/L_{\rm TIR}$  for
the stacked source by propagating the errors on both the stacked
X-ray and IR signal. We find that the stacked TBT-AGN lies $>3~\sigma$
above the expected range  for $L_X/L_{\rm TIR}$ for pure star-forming
galaxies, supporting our hypothesis that, on average, TBT-AGN harbor
AGN activity.  

\section{BPT Diagnostic: Stacking Analyses}
\label{stackBPT}


\begin{figure}[!htp]
\epsscale{1.2}
\plotone{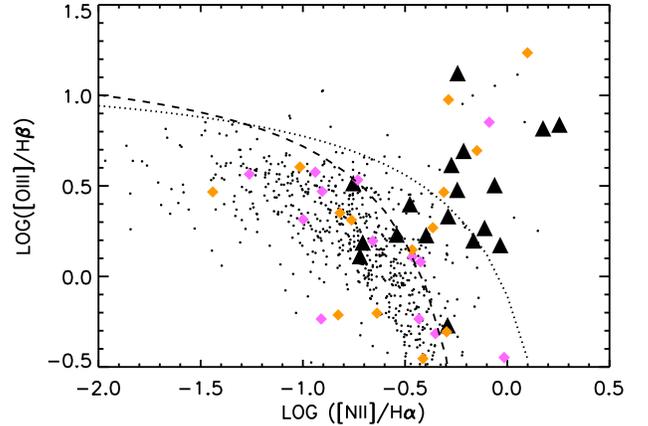}   
\caption{(a) BPT diagram for the GOODS-N/LH $z<0.5$ sources that have
  [OIII], H$\beta$, [NII], and H$\alpha$ fluxes with SNR~$>5$. 
  \emph{Large black triangles}---non-BLAGNs cross-listed in the
  \emph{Chandra} catalog with 
  $L_X>10^{42}$~\ergss. \emph{Orange diamonds}---non-BLAGNs
  cross-listed in the \emph{Chandra} catalog, but with
  $L_X<10^{42}$~\ergss. These sources are significantly detected in both the
  $0.5-2~\rm keV$ band and $2-8~\rm
  keV$ bands. \emph{Magenta diamonds}---non-BLAGNs cross-listed
  in the \emph{Chandra} catalog, but with $L_X<10^{42}$~\ergss. These sources
  are significantly detected
  in the $0.5-2~\rm keV$ band but not the $2-8~\rm keV$ band.
  \emph{Black circles}---remaining sources. The dashed
  curve shows the
  \citet{kauffmann03} empirical division between star-forming galaxies and
  AGNs. The dotted curve shows the \citet{kewley01} theoretical
  division.}
\label{goodsBPT}
\end{figure}

In Section \ref{TBT} we found that the majority of the SDSS BPT-comp lie within
the TBT-AGN regime. Here our primary goal is to do X-ray and IR
stacking analyses to investigate whether the BPT-comp signal, on
average, is dominated by AGN or star formation activity. 

Since the BPT diagram is restricted to sources with $z<0.5$, there are
too few GOODS-N BPT-comp for a robust stacking analysis
(specifically, there are only 22 GOODS-N BPT-comp). To increase our numbers,
here we also include the LH galaxy sample. Figure \ref{goodsBPT}
shows the BPT diagram for the 727 $z<0.5$ 
GOODS-N/LH galaxies that have
  [OIII], H$\beta$, [NII], and H$\alpha$ fluxes with SNR~$>5$. We have 
  excluded sources with broadened emission lines
  (FWHM~$>2000$~km~s$^{-1}$) from this analysis.  

As discussed in Section \ref{xray}, the BPT diagnostic does not match 100\%
with an X-ray selection of AGNs. In Figure \ref{goodsBPT} we see that
50\% (9/18) of the X-ray selected non-BLAGNs with
$L_X>10^{42}$~erg~s$^{-1}$~(black triangles) are BPT-AGN, 28\% (5/18)
are BPT-comp, and 22\% (4/18) are BPT-SF. 

In the following sections we do stacking analyses to
verify that the remaining $L_X < 10^{42}$~\ergss~BPT-AGN and BPT-SF are,
on average, AGN-dominated and SF-dominated,
respectively. Having tested our stacking analyses on these more
secure BPT categories, we apply them to our BPT-comp.

\subsection{X-ray Stacking}

Table \ref{table_stackBPT} provides the results from our X-ray stacking
analysis for the GOODS-N/LH BPT-SF, BPT-AGN, and BPT-comp. We followed
the same procedure as in Section \ref{GOODSxray}. We provide specifics
for each category in the following subsections. 

\begin{table*}[htp]
\small
\caption{BPT X-ray Stacking Analysis Results \label{table_stackBPT}}
\begin{tabular}{lccccccc}
\tableline\tableline
Category   & Total \#     &  \# used in stack    &  $f{0.5-2~\rm
keV}$\tablenotemark{a}  & $f_{2-8~\rm keV}$\tablenotemark{a} & $\langle
\Gamma_{\rm eff} \rangle$ & z\tablenotemark{b} & $\log L_{2-8~\rm keV}$
[erg~s$^{-1}$] \\
(1)             & (2)             & (3)                     & (4) & (5) & (6) & (7) & (8) \\
\tableline
BPT-SF & 605 &  448  &   $0.66 \pm 0.12$ ($5.5~\sigma$)  & $2.82 \pm 1.23$
($2.3~\sigma$)  & $1.5_{-0.3}^{+0.7}$\tablenotemark{c} & 0.41 & $40.20$\tablenotemark{c} \\
BPT-AGN\tablenotemark{d}    & 31   & 17  & $5.39 \pm 3.37$
($1.6~\sigma$)  & $39.4 \pm 22.9 $ ($1.8~\sigma$) &
$0.72_{-0.2}^{+1.5}$\tablenotemark{e} & 0.41 & $41.35$\tablenotemark{c} \\
 BPT-Comp    &  91   & 76  &  $1.71 \pm 0.33$ ($5.2~\sigma$)  & $8.92
\pm 2.23$ ($4.0~\sigma$)  &  $1.0_{-0.4}^{+0.4}$ & 0.41 & 40.70 \\
\tableline\tableline
\end{tabular}
\tiny\\
$^a$In units of $10^{-17}$~erg~cm$^{-2}$~s$^{-1}$.\\
\vspace{-0.05in}$^b$Volume-weighted redshift for the stacked source.\\
\vspace{-0.05in}$^c$We caution that this is based on a low
significance $2-8~\rm keV$ signal.\\
\vspace{-0.05in}$^d$Three out of four GOODS-N BPT-AGN and 11 out of 27 LH
  BPT-AGN are directly detected in their respective \emph{Chandra}
  image. \\
 \vspace{-0.05in}\hspace{0.06in}Stacking the remaining 17 GOODS-N/LH X-ray undetected
  BPT-AGN provides poor statistics for the stacking analysis.\\
\vspace{-0.05in}$^e$We caution that this is
  based on a low significance $0.5-2~\rm keV$ and $2-8~\rm keV$ signal.
\normalsize
\end{table*}

\subsubsection{BPT-SF}

There are 605 GOODS-N/LH BPT-SF. Twenty-one are directly X-ray
detected. Four of these are obvious X-ray selected AGNs with 
$L_X>10^{42}$~\ergss. Of the 17 X-ray detected sources with
$L_X<10^{42}$~\ergss, ten are only detected in the $0.5-2~\rm keV$
band. Using the $2-8~\rm keV$ flux limit for their \emph{Chandra} image (see Table
\ref{table_fluxlim}), we determine an upper limit to their
hardness ratios and find that all ten are X-ray soft, with $\Gamma >
1.7$. Of the seven remaining sources that are detected in both bands,
three are X-ray soft, with $\Gamma > 1.7$. Therefore, the majority
(13/17) of the X-ray detected, $L_X<10^{42}$~\ergss~sources are X-ray soft, with
$\Gamma>1.7$.

After excluding these known X-ray sources, as well as coincidental contaminants, we
use the STACKFAST program to determine the average X-ray signal in the
remaining 448 BPT-SF. The stacked source is
significantly detected in the $0.5-2~\rm keV$ band ($5.5~\sigma$) but
not in the $2-8~\rm keV$ band ($2.3~\sigma$). We find a $\langle
\Gamma \rangle = 1.5_{-0.3}^{+0.7}$, although we caution 
that this is based on a low-significance $2-8~\rm keV$ signal. Given the large
uncertainties, the stacked source could be X-ray soft ($\Gamma > 1.7$). 

Our MC simulations (see Section \ref{montecarlo} for details of the
procedure) yield a 2\%
probability of randomly generating the observed $0.5-2~\rm keV$ flux for our
stacked BPT-SF and a 96\% probability of generating the
observed $2-8~\rm keV$ flux. Our MC analysis confirms that these BPT-SF
clearly have an 
excess of $0.5-2~\rm keV$  counts, well above those we obtain
randomly. The $2-8~\rm keV$ signal is within the noise. 

\subsubsection{BPT-AGN}

There are 31 GOODS-N/LH BPT-AGN. Fourteen are directly X-ray
detected. Nine of these are obvious X-ray selected AGNs with 
$L_X>10^{42}$~\ergss. Of the five X-ray detected sources with
$L_X<10^{42}$~\ergss, one is only detected in the $0.5-2~\rm keV$
band. Using the $2-8~\rm keV$ flux limit for its \emph{Chandra} image (see Table
\ref{table_fluxlim}), we determine an upper limit to its
hardness ratio and find that it is X-ray soft, with $\Gamma >
1.7$. The four remaining sources that are detected in both bands are
X-ray hard, with $\Gamma <1.4$.

After excluding these known X-ray sources, as well as coincidental contaminants, we
use the STACKFAST program to determine the average X-ray signal in the
remaining 17 BPT-AGN. Given the small number of sources in this
stacking analysis, it is not surprising that the stacked source is neither significantly
detected in the $0.5-2~\rm keV$ band ($1.6~\sigma$) nor in the $2-8~\rm
keV$ band ($1.8~\sigma$).  We find a $\langle \Gamma \rangle =
0.72_{-0.2}^{+1.5}$, although we caution that this is based on a
low-significance $0.5-2~\rm keV$ and $2-8~\rm keV$ signal.  Given the large
uncertainties, this stacked source could be X-ray hard ($\Gamma < 1.4$). 

Our MC simulations (see Section \ref{montecarlo} for details of the
procedure) yield a $\sim 98$\% 
probability of randomly generating the observed $0.5-2~\rm keV$ and
$2-8~\rm keV$ fluxes. Our MC analysis confirms that the $0.5-2~\rm keV$
and $2-8~\rm keV$ signals are within the noise. 

\subsubsection{BPT-comp}

There are 91 GOODS-N/LH BPT-comp. Eleven are directly X-ray
detected. Five of these are obvious X-ray selected AGNs with
$L_X>10^{42}$~\ergss. Of the six X-ray detected sources with
$L_X<10^{42}$~\ergss, three are only detected in the $0.5-2~\rm keV$
band. Using the $2-8~\rm keV$ flux limit for their
\emph{Chandra} image (see Table \ref{table_fluxlim}), we determine an
upper limit to their hardness ratios and find that all three are X-ray
soft, with $\Gamma > 1.7$. The remaining three sources are detected in
both bands and are X-ray hard, with $\Gamma < 1.4$. 

After excluding known X-ray sources, as well as coincidental contaminants, we
use the STACKFAST program to determine the average X-ray signal in the
remaining 76 BPT-comp. The stacked source is significantly detected in
both bands  ($5.2~\sigma$ in the $0.5-2~\rm
keV$ band and $4.0~\sigma$ in the $2-8~\rm keV$ band). We find a
$\langle \Gamma_{\rm eff} \rangle =1.0_{-0.4}^{+0.4}$, i.e., the source is X-ray hard.  

Our MC simulations for
the BPT-comp yield a 5\% probability that we would
see the observed $0.5-2~\rm keV$ flux and a 2\% probability
we would see the observed $2-8~\rm keV$ flux, confirming that
we are recovering a real signal in both bands. Fewer than $1$\% of our
MC simulations for the BPT-comp result in both the $0.5-2~\rm keV$ and
$2-8~\rm keV$ bands detected at a $3~\sigma$ level or greater. In
other words, there is a $<1$\% 
false-alarm rate for our stacked BPT-comp signal. Thus, the
observed X-ray hardness in our BPT-comp appears to be a reliable result. 

\subsection{IR Stacking}

\begin{figure}[!htp]
\epsscale{1.2}
\plotone{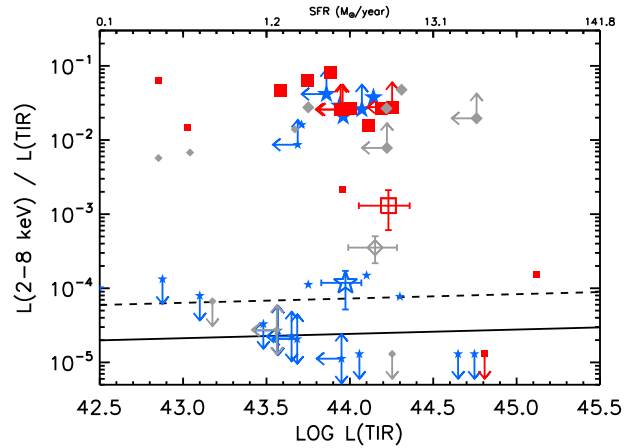}  
\caption{$L_{2-8~\rm keV}/L_{TIR}$ versus $\log L_{TIR}$ for our $z<0.5$
 GOODS-N/LH galaxies that have [OIII], H$\beta$, [NII], and H$\alpha$
 fluxes with SNR~$>5$. \emph{Red filled squares} -- directly X-ray
 detected BPT-AGN. \emph{Blue filled stars} -- directly X-ray detected 
  BPT-SF. \emph{Gray
    filled diamonds} -- directly X-ray detected BPT-comp.  Large (small)
  symbols designate sources with 
  $L_X>10^{42}$~erg~s$^{-1}$ ($L_X<10^{42}$~erg~s$^{-1}$). Downward
  pointing arrows indicate 
  sources undetected in the $2-8~\rm keV$ band (but detected in the
  $0.5-2~\rm keV$ band) that are
  assigned the $2-8~\rm keV$ flux limit (see Table
  \ref{table_fluxlim}). Upward and leftward pointing 
  arrows indicate sources undetected in the \emph{Spitzer} $24\mu$m
  image that are assigned the $24\mu$m flux limit  (see Table
  \ref{table_fluxlim}). \emph{Blue open star} -- stacked
  BPT-SF, excluding directly X-ray detected sources. \emph{Red open
    square} -- stacked BPT-AGN, excluding directly X-ray detected
  sources. \emph{Gray open
    diamond} -- stacked BPT-comp, excluding directly X-ray detected
  sources. \emph{Solid line} -- expected ratio for the HMXB population in 
  a galaxy with the given SFR (see top axis),
  according to \citet{persic04}. \emph{Dashed line} -- expected ratio
  for the overall star formation occurring in a galaxy with the given
  SFR, according to \citet{mineo11}.}
\label{LxIR_BPT}
\end{figure}

As discussed in Section \ref{IRstack}, the X-ray hard signal in our stacked
BPT-comp could be due either to HMXBs or to AGN activity with
obscuration. We follow the same procedure as in Section \ref{IRstack}
to distinguish between these scenarios using the IR. We determine the $1~\sigma$
error on $L_X/L_{\rm TIR}$  for 
the stacked BPT-comp (gray open diamond) by propagating the errors on
both the stacked 
X-ray and IR signal. Figure \ref{LxIR_BPT}  shows that the stacked BPT-comp lies
$>3~\sigma$ above the expected range for $L_X/L_{\rm TIR}$ for pure
star-forming galaxies (dashed line). This corroborates what we find for
the individually X-ray detected BPT-comp (gray filled diamonds), where the
majority (8/11) have
$L_X/L_{\rm TIR}$ values well above the expected range for pure
star-forming galaxies. These results support our TBT diagnostic inclusion of
the bulk of BPT-comp in the TBT-AGN regime. 

We also note that every one of the BPT-AGN, BPT-SF,
and BPT-comp with $L_X>10^{42}$~erg~s$^{-1}$~(large filled symbols)
have $L_X/L_{\rm TIR}$ values well above the expected range for pure star-forming
galaxies. This further confirms that the signal in these X-ray selected
AGNs is dominated by AGN activity. An optical classification as BPT-SF
is inaccurate. 

The stacked BPT-AGN (red open square in Figure \ref{LxIR_BPT}) was not
signitificantly detected in either the $0.5-2~\rm keV$ or $2-8~\rm
keV$ bands ($1.6~\sigma$ and $1.8~\sigma$, respectively). Nonetheless, we
use the derived $L_X$ from the stacking analysis
(see Table \ref{table_stackBPT}) and determine the $1~\sigma$
error on $L_X/L_{\rm TIR}$  by propagating the errors on
both the stacked X-ray and IR signal. The stacked BPT-AGN $L_X/L_{\rm
  TIR}$ lies well above the expected range for pure star-forming
galaxies. This corroborates what we find for the individually X-ray
detected BPT-AGN with $L_X<10^{42}$~erg~s$^{-1}$~(small red filled
squares), where the majority (3/5) have $L_X/L_{\rm TIR}$ well above the
expected range for pure star-forming galaxies. The two with
$L_{\rm TIR}>10^{44.5}$ lie within the expected range for pure
star-forming galaxies. These two sources are an additional example of the BPT 
diagnostic potentially misidentifying sources -- in this case,
SF-dominated sources as BPT-AGN.   

The stacked BPT-SF
(blue open star in Figure \ref{LxIR_BPT}) was significantly detected in the
$0.5-2~\rm keV$ band ($5.5~\sigma$) but not in the $2-8~\rm keV$ band
($2.3~\sigma$). Nonetheless, we use the derived
$L_X$ from the stacking analysis (see Table \ref{table_stackBPT}) and
determine the $1~\sigma$ error on $L_X/L_{\rm TIR}$  by propagating the errors on
both the stacked X-ray and IR signal. The stacked
BPT-SF $L_X/L_{\rm TIR}$ is consistent with that expected for pure
star-forming galaxies. This corroborates what we find for
the individually X-ray detected BPT-SF with
$L_X<10^{42}$~erg~s$^{-1}$~(small blue filled stars), where the majority (12/17) have
$L_X/L_{\rm TIR}$ values consistent with the expected range for pure
star-forming galaxies. 


\section{Discussion -- Comparison with Alternative Diagnostics}
\label{disc}

As mentioned in the Introduction, \citet{lamareille10} investigate the
use of a `blue diagnostic' -- [OIII]/H$\beta$ versus [OII]/H$\beta$ --
to classify galaxies as
star-forming galaxies, AGNs, and composites. Although
\citet{lamareille10} separate out LINERs\footnote{The right wing of the BPT diagram
further subdivides into an upper and lower branch, with LINERs being
the sources in the
lower branch exhibiting lower [OIII] luminosities
\citep{heckman80,kauffmann03}. The nature of these sources (weak
AGNs versus `retired' galaxies dominated by old stellar populations with
relatively hard radiation fields) is still much debated
\citep{kewley06,stasinska08,cidfernandes10,cidfernandes11}.}, here we
include both AGNs
and LINERs in the BPT-AGN category, as in the rest of our
article. Since [OIII], [OII], and H$\beta$ lie at the blue end of the
spectrum, this diagnostic can be applied to galaxies with optical
spectra out to $z \sim 0.9$. This method provides a successful means
for creating a reliable sample of star-forming galaxies out to these
redshifts. Their SF-dominated regime (SFG in their
Table 1) encloses $>99$\% of the BPT-SF. Only $1.5$\% of the
sources in the SFG regime are BPT-AGN. The `blue diagnostic', however, is not
as effective in creating a reliable AGN selection. Their combined Sy2,
SF/Sy2, and LINER regimes identify 
$\sim94$\% of all the BPT-AGN. However, $\sim23$\% of the sources in
this combined regime are BPT-SF\footnote{If we remove the LINER
  regime from this analysis, we note that the Sy2 and SF/Sy2 regimes
  enclose 59\% and 26\% of the Seyferts, respectively. While $<3$\% of
  sources in the Sy2 regime are BPT-SF, 74\% of sources in the SF/Sy2
  regime are BPT-SF.}. 

A classification scheme based on
[NeIII]/[OII] (\citealt{stasinska06}; this work) complements this
`blue diagnostic' by correctly classifying galaxies in the \citet{lamareille10} SF/Sy2
category. \citet{marocco11} recently examined the location of the
\citet{lamareille10} SF/Sy2 galaxies in 
the \citet{stasinska06} DEW diagnostic -- [NeIII]/[OII] versus
$D_n[4000]$. They find that by applying an additional empirical cut in
the DEW diagnostic space to these SF/Sy2 galaxies, they are able
to correctly identify 99\% of the BPT-SF and 97\% of the Seyfert
2s. Overall, this combined diagnostic correctly
identifies 85\% of Seyfert 2s and 99\% of BPT-SF. 

Our TBT diagnostic, based on [NeIII]/[OII] versus rest-frame $g-z$
color (rather than $D_n[4000]$), results in minimal
overlap between the BPT-SF and BPT-AGN, with only $\sim
1.3$\% of the BPT-AGN lying within the TBT-SF regime and only
$2.8$\% of BPT-SF lying within the TBT-AGN regime. Likewise, of the
sources in the TBT-AGN regime, only 3.5\% are BPT-SF and of the
sources in the TBT-SF regime, only 1\% are BPT-AGN.

The \citet{juneau11} MEx diagnostic -- [OIII]/H$\beta$ versus
stellar mass -- also complements the `blue diagnostic' in its ability
to reliably identify AGNs. The MEx-AGN regime encloses 99\% of
BPT-AGN. Only 6\% of the sources in the MEx-AGN regime are
BPT-SF. Similarly, the \citet{yan11} CEx diagnostic -- [OIII]/H$\beta$
versus rest-frame $U-B$ color -- also
complements the `blue diagnostic', with the CEx-AGN regime enclosing
95.7\% of BPT-AGN. Only 1.9\% 
of the sources in the CEx-AGN regime are BPT-SF. 

An important difference between the MEx and CEx diagnostics and our TBT diagnostic 
is the classification of X-ray selected AGNs. While 8\% (8/101) of the
Juneau et al.~and 22\% (30/126) of the Yan et al.~X-ray selected  
$L_X>10^{42}$~erg~s$^{-1}$~AGNs lie in the MEx-SF and CEx-SF
regimes of their diagnostics, respectively, only 3\% (3/103) of our X-ray selected
AGNs lie within our TBT-SF regime (see Section \ref{xray}). This may
be a result of the higher ionization potential of the [NeIII]
line (63.45 eV) as compared with the [OIII] line (54.94 eV). The higher
ionization potential appears to foster a more reliable selection of
AGN-dominated galaxies, i.e., the weaker ionizing flux in star-forming galaxies
lessens their ability to excite [NeIII] as compared to [OIII]. 

We note that our approach does not address the issue of classifying
galaxies with very low equivalent width emission lines, where some of the lines are
too noisy for traditional line diagnostics to be used. See
\citet{cidfernandes10,cidfernandes11} for their 
discussion of the WHAN diagram --
$W_{\rm H \alpha}$ versus [NII]/H$\alpha$ -- and its ability to provide
a more comprehensive emission line classification of galaxies. 

\subsection{BPT-comp: AGN-dominated sources}

A critical difference between the \citet{lamareille10} and CEx diagrams and
our TBT diagnostic is the location of the BPT-comp (galaxies that fall between the
\citealt{kauffmann03} and \citealt{kewley01} divisions in the BPT
diagram). In the \citet{lamareille10} diagram, $\sim 83$\% of BPT-comp lie
within the SFG regime. In the combined \citet{lamareille10} and
\citet{marocco11} diagnostic,
$\sim60$\% of BPT-comp lie within the SFG+SFG/comp regimes. In the CEx
diagram, $\sim75$\% of BPT-comp lie within 
the CEx-SF regime. In our TBT diagnostic, on the other
hand, 65\% of the BPT-comp lie within the TBT-AGN regime, with
only 35\% in the TBT-SF regime.  Similarly, in the MEx diagnostic
\citep{juneau11}, only 17\% of BPT-comp lie within the MEx-SF regime. 

A number of optical emission-line studies have argued that the signal
in BPT-comp is dominated by star formation activity, rather than AGN activity.
The \citet{kewley01} upper boundary to the BPT-comp regime marks their
 theoretical prediction for galaxies whose contribution from AGN
 activity to the extreme ultraviolet ionizing radiation field just
 begins to exceed 50\%. According to this work, all sources to the
 lower-left of this boundary have their signal dominated by star formation
 activity. Similarly, \citet{stasinska06} use spectral 
 synthesis modeling to argue that the contribution from AGN
activity to the emission-line signal in BPT-comp is 20\% or
less. Furthermore, \citet{kewley06} note that BPT-comp lie in the same parameter
space as HII regions in the [OIII]/[OII] versus [OI]/H$\alpha$
diagnostic (as well as within the SF-dominated regime of the
[OIII]/H$\beta$ vs. [SII]/H$\alpha$ diagnostic). They argue that this
provides further support for the idea that the ionizing 
radiation field and ionization parameter in BPT-comp are
dominated by star formation activity. 

Here we argue that our TBT diagnostic's reliance on [NeIII], with its
higher ionization potential than [OIII], [NII], or [SII], 
leads to a more reliable identification of AGN-dominated sources. In Section
\ref{stackBPT} we tested whether the inclusion of the majority of
BPT-comp in our TBT-AGN regime was justified. Of the individually X-ray detected
BPT-comp, 70\% are X-ray hard with $L_X/L_{\rm TIR}$ ratios indicative
of dominance by AGN activity (see Figure \ref{stackBPT}). The stacked
signal from X-ray undetected BPT-comp is also X-ray hard and exhibits
an $L_X/L_{\rm TIR}$ ratio $>3~\sigma$ above the expected range for
pure star-forming galaxies. This supports our TBT diagnostic inclusion
of BPT-comp in the TBT-AGN regime and suggests that, on average, the
X-ray and optical signal in BPT-comp is dominated by AGN activity. 


\section{SUMMARY}
\label{summary}

We have shown that the TBT diagnostic -- rest-frame \gminz\ color versus
[NeIII]/[OII] -- reliably separates SDSS SF-dominated sources from
AGN-dominated sources, as classified according to the classic BPT
diagram. Because both [NeIII] and [OII] are located
in the blue end of the optical spectrum, we are able to classify
galaxies using this diagnostic out to $z=1.4$. The TBT diagnostic
provides a significant extension
in redshift compared to the BPT diagram (limited in its use
with optical spectra to $z<0.5$) and the more recent
[OIII]/H$\beta$-based diagnostics (limited to $z<0.9$ -- `blue
diagram', \citealt{lamareille10}, \citealt{marocco11}; CEx, \citealt{yan11}; MEx,
\citealt{juneau11}).  

We find that the TBT selection of AGNs matches well with
an X-ray selection of AGNs, with 97\% (100/103) of our OPTX X-ray
selected AGNs lying within the TBT-AGN regime. This suggests that the
TBT diagnostic is more reliable than the BPT diagnostic in identifying
X-ray selected AGNs, since the BPT diagnostic misidentifies $\sim20$\% of
of our OPTX X-ray selected AGNs as BPT-SF, i.e., as star-forming galaxies. This may
be a result of the higher ionization potential of the [NeIII]
line (63.45 eV) as compared with the [OIII] line (54.94 eV). The higher
ionization potential appears to foster a more reliable selection of
AGN-dominated galaxies, i.e., the weaker ionizing flux in star-forming galaxies
lessens their ability to excite [NeIII] as compared to [OIII]. 

We perform X-ray and IR stacking analyses of the GOODS-N sample of
galaxies with accompanying deep \emph{Chandra} imaging to verify the
dominance of AGN activity in our TBT-AGN and star formation activity in our
TBT-SF. We find that the TBT-AGN, on average, are X-ray hard 
with $L_X/L_{\rm TIR}>3~\sigma$ above the expected range for pure
star-forming galaxies. Their X-ray hardness and excess X-ray signal is
likely due to obscured or weak AGN activity. The TBT-SF, on the other
hand, are X-ray soft with $L_X/L_{\rm TIR}$ consistent with pure
star-forming galaxies.  

We perform the same stacking analyses on the BPT categories in
order to confirm the selection of the majority of BPT-comp as TBT-AGN. As
expected, the BPT-SF are X-ray soft with $L_X/L_{\rm TIR}$ consistent
with pure star-forming galaxies and the BPT-AGN are X-ray hard with
$L_X/L_{\rm TIR}>3~\sigma$ above the expected range for pure
star-forming galaxies. Of the individually X-ray detected
BPT-comp, 70\% are X-ray hard with $L_X/L_{\rm TIR}$ ratios indicative
of dominance by AGN activity. Our stacked
BPT-comp is significantly  detected in both the $0.5-2~\rm keV$ and
$2-8~\rm keV$ bands and is 
X-ray hard, with $\langle \Gamma_{\rm eff} \rangle =
1.0_{-0.4}^{+0.4}$. Furthermore, the stacked BPT-comp $L_X/L_{\rm
  TIR}$ is $>3~\sigma$ above the expected range for pure star-forming
galaxies. These findings support our TBT diagnostic inclusion of BPT-comp
in the TBT-AGN regime. 

The BPT-comp (individual and stacked) properties suggest that, on
average, both their X-ray and optical signal is dominated by 
obscured or weak AGN activity. This is in contrast to claims by
previous optical emission-line studies that the signal in BPT-comp is
dominated by star formation activity. Therefore, we
recommend that groups carefully consider the impact of
excluding or including BPT-comp on the interpretation of their
results. For example, for studies involving
determining the bolometric contribution from AGN activity or the role
of AGN activity in galaxy evolution, we advise
maximal inclusiveness. Since BPT-comp comprise a significant
percentage of the overall emission-line galaxy population (20\% of the
SDSS DR8 sample), inclusion of the
BPT-comp would provide a more comprehensive picture of the true impact
of AGN activity in these studies. 

On the other hand, for metallicity studies of star-forming galaxies, we advise
maximal conservativeness \citep[e.g.,][]{tremonti04}. Emission lines
like [OIII] are boosted by AGN
activity and can masquerade as indicators of low metallicity, leading
to the misinterpretation of results. Therefore, in this case, it is
optimal to use a diagnostic that reliably excludes all AGNs and
AGN/SF composites. 

\acknowledgements
The authors thank the referee for comments and suggestions which
helped to improve this manuscript. We thank Ryan Hickox for helpful
discussions and training in the use 
of his STACKFAST X-ray stacking code. We gratefully 
acknowledge support from NSF grant AST 0708793, the University of
 Wisconsin Research Committee with funds granted by the Wisconsin
 Alumni Research Foundation, and the David and Lucile
 Packard Foundation (A.~J.~B.). LT acknowledges support through a
 CIERA postdoctoral fellowship.  

Funding for SDSS-III has been provided by the Alfred P. Sloan
Foundation, the Participating Institutions, the National Science
Foundation, and the U.S. Department of Energy Office of Science. The
SDSS-III web site is http://www.sdss3.org/. 

SDSS-III is managed by the Astrophysical Research Consortium for the
Participating Institutions of the SDSS-III Collaboration including the
University of Arizona, the Brazilian Participation Group, Brookhaven
National Laboratory, University of Cambridge, University of Florida,
the French Participation Group, the German Participation Group, the
Instituto de Astrofisica de Canarias, the Michigan State/Notre
Dame/JINA Participation Group, Johns Hopkins University, Lawrence
Berkeley National Laboratory, Max Planck Institute for Astrophysics,
New Mexico State University, New York University, Ohio State
University, Pennsylvania State University, University of Portsmouth,
Princeton University, the Spanish Participation Group, University of
Tokyo, University of Utah, Vanderbilt University, University of
Virginia, University of Washington, and Yale University. 

We wish to recognize and acknowledge the very significant cultural
role and reverence that the summit of Mauna Kea has always had within
the indigenous Hawaiian community. We are most fortunate to have the
opportunity to conduct observations from this mountain. 

\bibliographystyle{apj}  

\end{document}